\documentclass[apj]{emulateapj}

\usepackage[usenames]{color}
\usepackage{epsfig,graphicx}
\usepackage{natbib}

\newcommand{\rsun}{$R_{\odot}$}
\newcommand{\kms}{km~s$^{-1}$}

\newcommand{\lya}{Ly$\alpha$}
\newcommand{\lyb}{Ly$\beta$}

\newcommand{\hi}{H~{\sc{i}}}

\newcommand{\siiii}{Si~{\sc{iii}}}
\newcommand{\ciii}{C~{\sc{iii}}}
\newcommand{\water}{\mathrm H_2 \mathrm O}
\newcommand{\oh}{\mathrm{OH}}
\newcommand{\oo}{\mathrm{O}}

\shorttitle{Sun--Grazing Comet C/2002 S2}
\shortauthors{Giordano et al.}

\begin{document}

\title{Probing the Solar Wind Acceleration Region with the Sun--Grazing Comet C/2002 S2} 
\author{S. Giordano\altaffilmark{1}, J. C. Raymond\altaffilmark{2}, P. Lamy\altaffilmark{3},  
M. Uzzo\altaffilmark{4} and D. Dobrzycka\altaffilmark{5}}
\altaffiltext{1}{INAF--Osservatorio Astrofisico di Torino, via Osservatorio 20, 10025 Pino Torinese, Italy}
\altaffiltext{2}{Harvard--Smithsonian Center for Astrophysics, 60 Garden St., Cambridge, MA 02138, USA}
\altaffiltext{3}{Laboratoire d'Astrophysique de Marseille, 38 rue Fr\'ed\'eric Joliot--Curie, 13388 Marseille cedex 13, France.}
\altaffiltext{4}{Computer Science Corporation, 3700 San Martin Drive, Baltimore, MD 21218 USA}
\altaffiltext{5}{European Southern Observatory, Karl Schwarzschild Strasse 2, 85748 Garching (Germany)}

\begin{abstract}

Comet C/2002 S2, a member of the Kreutz family of Sungrazing comets, was discovered in 
white light images of the SOHO/LASCO coronagraph on 2002 September 18 and observed in \hi\, \lya\, 
emission by the SOHO/UVCS instrument at four different heights as it approached the Sun.
The \hi\, \lya\, line profiles detected by UVCS are analyzed to determine the spectral parameters: 
line intensity, width and Doppler shift with respect to the coronal background. 
Two dimensional comet images of these parameters are reconstructed at the different heights.
A novel aspect of the observations of this sungrazing comet data is that, whereas the emission from 
the most of the tail is blue--shifted, that along one edge of the tail is red--shifted.  We attribute
these shifts to a combination of solar wind speed and interaction with the magnetic field.
In order to use the comet to probe the density, temperature and speed of the corona and
solar wind through which it passes, as well as to determine the outgassing rate of the comet,
we develop a Monte Carlo simulation of the \hi\, \lya\, emission of a comet moving through a coronal plasma. 
From the outgassing rate, we estimate a nucleus diameter of about 9 meters.
This rate steadily increases as the comet approaches the Sun while the optical brightness decreases by
more than a factor of ten and suddenly recovers. This indicates that the optical brightness
is determined by the lifetimes of the grains, sodium atoms and molecules produced by the comet.

\end{abstract}


\section{Introduction}

Over 2000 comets have been discovered by the LASCO coronagraphs aboard the SOHO spacecraft, 
most of them members of the Kreutz family of sungrazing comets.  
These comets follow similar orbits, with perihelion close to the surface of the Sun, so few survive the encounter,
a notable exception being Comet Lovejoy (C/2011 W3) \citep{mccauley13, downs13, raymond14}.  
The Kreutz family is believed to come from the breakup of a single progenitor at least 1700 years in the past 
 \citep{sek04A,mar05}, and close pairs of comets indicate that breakup occurs 
throughout the orbit \citep{sek07A}.

The LASCO observations show a consistent pattern of brightness as a function of distance from the Sun.  
The comets increase rapidly in brightness until they reach about 12~\rsun, at which point they begin to fade
rapidly, presumably because the dust that scatters visible light begins to sublimate rapidly \citep{bie02}.  
The brightest ones are seen to level off at a lower brightness at smaller radii, though the brightness fluctuates.  
This may be a sign that the nucleus is breaking up.

The UVCS instrument aboard SOHO has obtained ultraviolet spectra of a number of Kreutz family comets.  
Water outgassing from the comet is rapidly photodissociated.  
The resulting hydrogen atoms can scatter \lya\, photons from the solar disk, but their motion toward the Sun 
Doppler shifts the scattering cross section profile away from the solar emission line 
profile (Doppler dimming, also known as the Swings effect).  
Therefore, most of the observed \lya\, comes from H atoms that have gone through charge transfer with 
the ambient coronal or solar wind protons, and they have a velocity distribution similar to that of the proton thermal 
distribution.  We refer to the neutral populations before and after charge transfer as first and second
generation hydrogen atoms, respectively.

The UVCS spectra have been used to determine the solar wind speed at 6.8 \rsun\, in a coronal hole from the line 
width and the conditions for the comet bow shock \citep{ray98B}.  
They imply nucleus sizes on the order of 10 m \citep{ray98B,uzz02,bem05} to hundreds of meters \citep{mccauley13}.
Increases in the apparent size of the nucleus imply breakup, providing an estimate of the tensile strength of the 
nucleus \citep{uzz02}.  The time scale for fading of the \lya\, brightness makes it possible to determine the 
coronal density without integration along a line--of--sight  \citep{uzz01}.  A persistent \lya\, signature can 
be interpreted in terms of a refractory 
population of grains \citep{kim02,bem05}. 
UVCS observations of  \ciii\,  and \siiii\, lines from a bright sungrazer indicate an overabundance of Si compared 
to C in the cometary dust \citep{cia10}.

The comet C/2002 S2 is a bright member of the Kreutz family discovered on 18 September, 2002.  
It reached an apparent V magnitude of 3.3, then faded suddenly by over an order of magnitude in 
the optical when it reached a height of 5.7~\rsun, then recovered just as rapidly.  
Meanwhile, the \lya\,  brightness observed by UVCS increased steadily as the optical faded.  
In this comet we detect \lya\,  from both pre-- and post--charge transfer
hydrogen atoms, providing additional constraints on the outgassing
rate and the coronal parameters.  

The most remarkable feature of the observations is a substantial blue--shift of the northern
part of the tail and red--shift of the southern part.  
This requires some mechanism to break the symmetry of the coronal velocity distribution or the charge transfer process.  
We suggest that the blue--shift is the line--of--sight component of the solar wind, while the
red--shift results from pickup ion process similar to that studied in Comet Lovejoy by \cite{raymond14}. 
When an atom is ionized it becomes a pickup ion with velocity components parallel and perpendicular
to the magnetic field, and in this case the parallel component is away from the Earth.  
Subsequent charge transfer events with other neutrals produce a red--shifted population of hydrogen atoms moving 
in the direction of the magnetic field.

We construct a three--dimensional time--dependent Monte Carlo model of the kinematics (trajectory) of the 
outgassed neutral hydrogen, taking into account the ionization and charge transfer processes and use it to 
infer the outgassing rates and coronal parameters, such as outflow wind velocity, electron density and 
proton temperature.  While most remote sensing observations provide only line--of--sight integrated
quantities and averages, the comet allows us to probe individual points along the comet's path.
Section~2 describes the observations and comet kinematics. 
In Section~3 we present an overview of the physical processes involved in the \lya\, emission, 
Section~4 describes the Monte Carlo simulations and comparison with observation, and 
Section~5 discusses the derived comet and coronal parameters and the discrepancies between model and observation.

\section{Observations}

\subsection{LASCO Observations and Comet Kinematics}

The comet C/2002 S2 was discovered in LASCO data on 18 September, 2002 at a heliocentric distance 
larger that 16~\rsun\, and it was followed down to about 3.0~\rsun, where it progressively disappeared
and apparently sublimated before perihelion. 
The two coronagraphs involved in the observation are C2 and C3. 
A clear filter was used for C3 observation, with a nominal bandpass of 4000 to 8500~\AA,
while C2 observes with an orange filter, which selects the bandpass from 5400 to 6400~\AA.
For a detailed description of LASCO system, see \citet{bru95}.

\begin{figure}
  \centerline{\epsfig{file=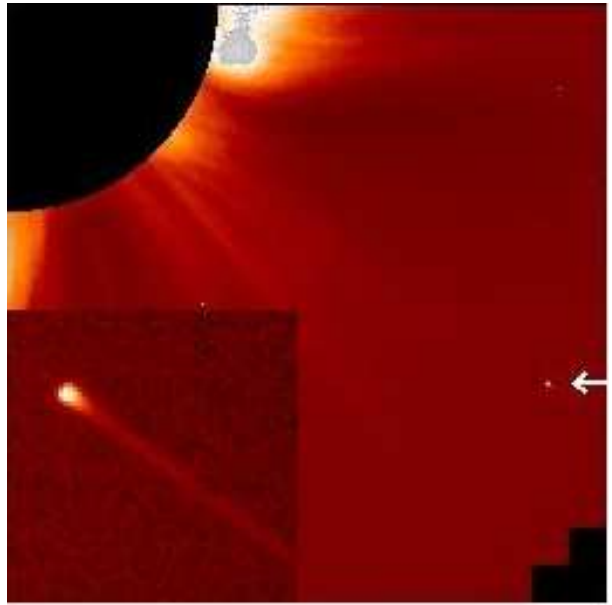,width=8.cm,angle=0}}
  \caption{LASCO C2 comet  C/2002 S2 observation on September 18, 2002 at 21:30 UT. 
  In the bottom left inset a zoomed image of the comet is shown.}
\end{figure}

The comet approaches the Sun from the southwest at a position angle of about 236$^{\circ}$ 
counterclockwise from north pole. In Figure~1 we show the LASCO C2 image taken on
September 18, 2002 at 21:30 UT, when the comet was at the heliocentric distance of about 6.9~\rsun\, projected 
into the plane of the sky; in the same figure we show a detail of the comet image where the comet tail is 
clearly visible.  The visible length of the tail is real because in the 25 second exposure time, 
the comet, moving in the plane of the sky at a velocity of about 200~\kms\, covers a spatial region smaller than 
the C2 pixel size, which is 11.9 arcsec.
The orbital parameters released in the Minor Planet Electronic Circular issued on  September 18, 2002 
\citep{uzz02} made it possible to adjust the UVCS pointing in order to detect the comet's ultraviolet 
emission and a subsequent more precise comet ephemeris computation MPEC 2002--S36 (see Table~1) 
permits a more accurate comet orbit  computation and therefore the determination of the position and the 
kinematic parameters of the comet.

\begin{figure}
	\centering
	\epsfig{file=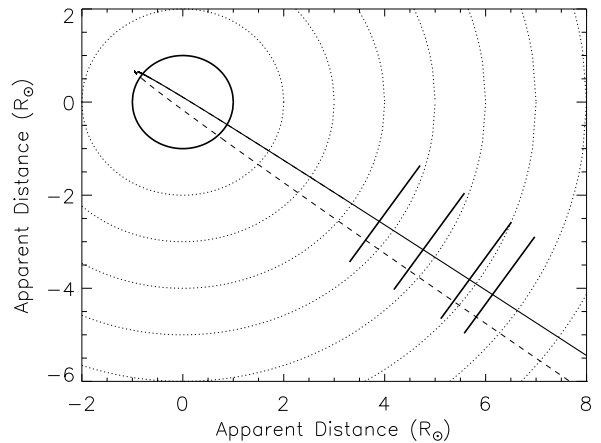,width=6.5cm,angle=90}
	\caption{Comet C/2002 S2 trajectory in the plane of the sky as computed from the orbital parameters. 
	We plot both the pre--perihelion (solid line) and the expected post--perihelion (dashed line). 
	The UVCS slit positions at the observed heliocentric distances are also shown by the solid lines perpendicular
	to the trajectory.}
\end{figure}

\begin{table}
\caption {Comet C/2002 S2 orbital elements}
\centerline{
\begin{tabular}{|c|c|l|}
\hline
{\it Element} & {\it Value} & {\it Description} \\
\hline
$T_{pass}$      & 2002~Sept.~19.12~TT       & Time of perihelion passage     \\
$q$             & 0.0053~AU                 & Perihelion distance             \\
$Peri=\omega$   & 81.07$^{\circ}$           &        Argument of perihelion   \\
$Node=\Omega$   & 3.76$^{\circ}$            & Longitude of the ascending node \\
$Incl=i$        & 144.31$^{\circ}$          & Inclination                     \\
$\epsilon$      & 1.0                       & Eccentricity                    \\
\hline
\end{tabular}
}
\end{table}

The comet orbit projected into the plane of the sky is shown in Figure~2 with the UVCS slit positions superimposed.
Table~2 shows the comet position and kinematic parameters, computed from ephemeris, 
at the time of the first UVCS comet detection at each observed height, $t_{enter}$, given by the average 
time between the beginning and the end of the UVCS exposure which first detected the comet in \hi\, \lya.
The actual heliocentric distance of the comet from the Sun is given by $r$, and $\rho_{obs}$ is the comet--Sun 
distance in the plane of the sky, that is the observed height.
The comet is moving toward the Sun with a phase angle,  $\alpha$, ranging from $\approx$30 to 40$^{\circ}$, 
which is the angular distance of the comet from the plane of the sky toward the Earth. 
We verify that the expected positions projected into the plane of the sky from computation agree well with 
the observed positions by LASCO and UVCS.
In Table~2 we also show for each height the number of UVCS exposures detecting comet and tail in column 
$N_{exp}$ and the peak of the measured \hi\, \lya\, flux, F$_\alpha$, in $\rm photons~cm^{-2}~s^{-1}$ 
(see Section~2.2).

\begin{table}[h]
\caption {Comet C/2002 S2 position and kinematic parameters at the time of UVCS comet observations}
\centerline{
\begin{tabular}{|c|c|c|c|c|c|c|c|c|}
\hline
$t_{enter}$ & $N_{exp}$ &  $r$    & $\rho_{obs}$  & $\alpha$  & $V_{r}^{\#}$ & $V_{los}^{\#}$   & $V_{pos}^{\#}$   &F${_\alpha}^{*}$ \\ 
      (UT)     &   &(\rsun)  & (\rsun) & (deg)        &    &         &            &     \\
\hline
21:05:21 &  10 & 8.54 & 7.40 & 30 & 197 &  31 &  -209 & 	~~45 \\
21:32:31 &  24 & 8.02 & 6.84 & 31 & 202 &  35 &  -215 & 	192 \\
22:29:39 &  21 & 6.97 & 5.72 & 35 & 214 &  45 &  -230 & 239 \\
23:18:26 &  28 & 5.99 & 4.66 & 39 & 227 &  57 &  -246 & 474 \\
\hline
\end{tabular}
}
\end{table}
\noindent
\# (\kms) \\
$*$ ($\rm ph~cm^{-2}~s^{-1}$) \\

The comet kinematic parameters are given by the velocity components toward the Sun, $V_{r}$,  
along the line--of--sight, $V_{los}$,  and in the plane of the sky, $V_{pos}$.
The comet light curve from LASCO C2 and C3 observations is shown in Figure~3.
The apparent magnitude of the comet coma is plotted as a function of time, 
thus with decreasing heliocentric distance.  The brightness differs somewhat
from the values shown in Figure 2 of \cite{kni10} because we have not corrected the observed magnitudes
for the phase angle, but both light curves show the dramatic minimum near 6 \rsun.  We note that 
the coma brightness increases as the comet 
approaches the Sun and it reaches a peak at $\sim$~12~\rsun\, then it decreases up to $\sim$~5.7~\rsun\,
at this point the brightness shows a sudden increase which continues until the comet is observed at the closest
heliocentric distance of  $\sim$~3.0~\rsun.
The increasing comet brightness above 12 \rsun\, is a result
of increasing solar flux and an increasing outgassing rate. After the peak the sublimation rate of the 
coma dust is higher than its production rate, giving the decreasing brightness observed for all the 
Kreutz sungrazing comets \citep{bie02,kni10}. 
Finally, the increase at lower distances might be related to fragmentation events at the distance of minimum of the 
brightness ($\sim$~5.7~\rsun) or farther from the Sun \citep{bem05}.
The apparent \hi\, \lya\, magnitude from UVCS data, superposed on the visible magnitude from LASCO
in Figure~3 is discussed in Section~2.2.

\begin{figure}[h]
\centering
\epsfig{file=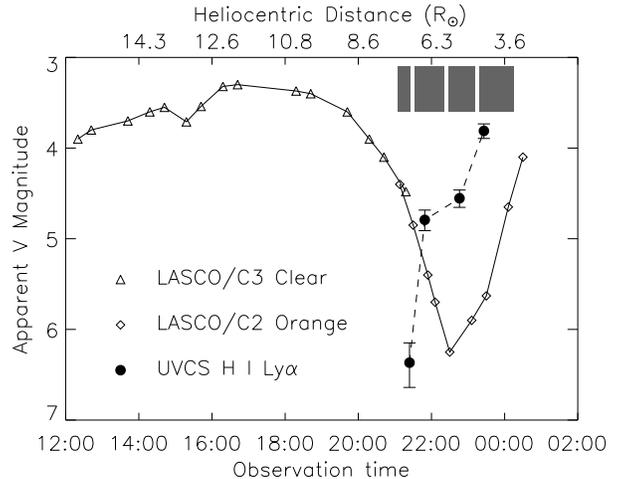,width=9cm}
\caption{Apparent V magnitude of Comet C/2002 S2 determined from LASCO C2 and C3.  
The black dots represent the apparent magnitudes derived from the UVCS observations of the \hi\, \lya\, line
arbitrarily scaled to the LASCO apparent magnitude range.}
\end{figure}

\subsection{UVCS Observations and Spectral Data Analysis}

UVCS observed Comet C/2002 S2 on 2002 September, 18 
between 20:36 UT and 00:15 UT of the next day. The instrument roll to catch the comet was 236$^{\circ}$ 
counterclockwise from north pole and the heliocentric distance ranges from 8.10~\rsun\, to 4.66~\rsun.
After an initial 120 second exposure at 8.10~\rsun, where a faint signal from the comet was observed, 
the data were acquired at 4 lower heights: 7.40, 6.84, 5.72, and 4.66~\rsun\, in a series of 120 $s$ exposures. 
The number of exposures, $N_{exp}$, at each height is given in Table~2.
The observations were designed so the coverage at each height includes some exposures before and/or after 
the comet enters the UVCS slit to provide information about the coronal and interplanetary \hi\, \lya\, background.
The UV spectral and spatial binning were 0.183~\AA\, (2~pixels) and 21$''$ (3~pixels) respectively. 
The slit width was 150~$\mu$m, which corresponds to an integration region of 42$''$ in the direction of the comet path.
In the UV spectral range covered (975 to 1223~\AA\, to 1039.35~\AA)  the comet was detectable only in the 
\hi~\lya\, 1215.67~\AA\, line.
The UVCS position angle,  236$^{\circ}$ from solar north, was successfully chosen so that the comet crossed 
near the center of the slit and the comet path is perpendicular to the slit length, as shown also in Figure~2.
The temporal sequence of \hi\, \lya\, spectral images acquired when the comet crossed
the lowest observed height, i.e. $\rho_{obs}$=4.66\rsun, is reported in Figure~4.  The emission 
increases as the comet nucleus enters the slit (first three exposures), and then
fades as the hydrogen is ionized away.
We can see also the increasing
size of the \hi\, comet cloud due to the thermalization of the outgassed comet hydrogen with the coronal protons. 
The uncertainty in the determination of comet position with the UVCS data, besides the calibration uncertainties, 
is due to the slit width and the comet movement through the slit during the integration time with a velocity range from
$\approx$209 to $\approx$246 km $\rm s^{-1}$ at the different observed heights,  
therefore it is estimated to be at least 0.04~\rsun.

\begin{figure}[h]
\centering
\epsfig{file=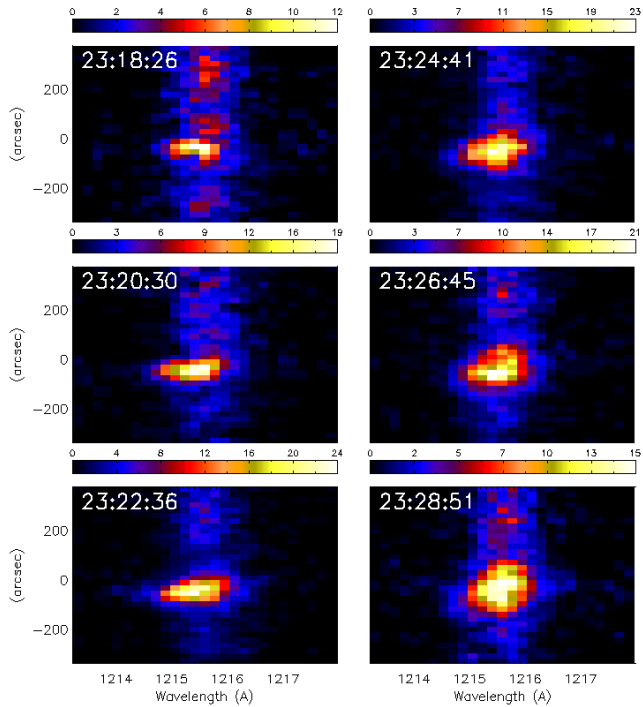,width=8.5cm}
\caption{Sample temporal sequence of \hi\, \lya\, spectral images at the time of comet crossing UVCS slit 
	at $\rho_{obs}$=4.66~\rsun.
	The color bars show the total count per pixel collected in 120 second exposures. 
	In each panel the x--axis represents the wavelength and the y--axis the position along the slit.}
\label{fig:spectra}
\end{figure}

Before the comet observation, the UVCS daily synoptic program, which lasted about 12 hours, was performed by scanning 
the 360$^{\circ}$ corona at eight polar angles and several heliocentric heights. 
In order to remove possible instrumental effects in the comet spectra, we perform an evaluation of the 
spatial flat field along the slit, by combining all the \hi\, \lya\, counts in the spectral direction for all the synoptic 
scans performed on 2002 September, 18. Then we normalized the profile along the slit by a smoothed profile to obtain
the flat field which is then applied to the spectral data.
The analysis of spectral data is performed after the subtraction, at each height, of the off--line background,
evaluated in a detector region where no lines are expected, and of the background from the exposures  
before the comet. In this way we remove the background emission due to 
coronal, stray light and interplanetary \hi\, \lya.

The background emission can be used to evaluate the coronal proton temperatures before the 
comet crossing.  Stray light and interplanetary emission are very narrow with respect to the coronal line, 
so the background values have to be considered lower limits to the coronal proton temperatures. 
Thus, by taking into account the instrumental broadening, the coronal kinetic temperature can be estimated 
as $T_{k}~\sim 1.1~\times~10^{6}~K$ at $\rho_{obs}$=4.66~\rsun\, decreasing to 
$T_{k}~\sim 6.4~\times~10^{5}~K$ at $\rho_{obs}$=6.84~\rsun;
we note that these are average temperatures along the lines of sight, so they do not necessarily apply to 
the regions the comet crossed.

The \hi\, \lya\, comet light curves, that is the intensity as a function of time, are computed by integrating
the spectra over $\pm$300" along the slit around the comet center and fitting them by a gaussian profile
convolved with the instrumental profile. The light curves are shown in Figure~5 for the 4 observed heights.
We see that, at the lowest height ($\rho_{obs}$=4.66~\rsun, black curve) the emission increases quite rapidly and 
slowly fades to pre--comet values, while at higher distances the growth is more gradual. 
Finally, at the largest observed distance ($\rho_{obs}$=7.40~\rsun, green curve), the signal is very noisy and
useless for spectral analysis.
We point out that, at the time of the first contact, the real signal could be larger than observed because the comet 
is not yet completely into the UVCS slit. 

\begin{figure}[h]
\centering
\epsfig{file=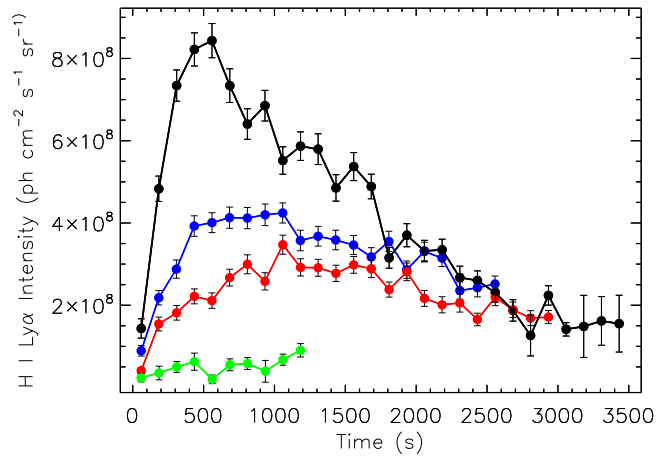,height=8.cm,angle=90}
  \caption{Observed \hi\, \lya\, intensity (photons $\rm cm^{-2}~s^{-1}~sr^{-1}$) as a function of time 
  	      integrated over a region of $\pm$300" centered on the comet axis from observations at  
   	      $\rho_{obs}$=4.66~\rsun\, (black), 
               $\rho_{obs}$=5.72~\rsun\, (blue), 
               $\rho_{obs}$=6.84~\rsun\, (red) and 
               $\rho_{obs}$=7.40~\rsun\, (green).}
  \label{fig:lightcurves}
\end{figure}

From gaussian fitting of the cometary spectra integrated along the slit we also determine the centroid 
of \hi\, \lya\, line with respect to the background corona.
Whereas at the 3 higher heights the values are comparable, within the uncertainties, to the pre--corona 
emission, at 4.66~\rsun\, a clear trend from a 100 $\rm km~s^{-1}$ blue--shift toward the velocity of the 
background corona is found, as shown in the left panel of Figure~6.  We note that the spectrum in the first 
exposure when the comet is coming into the UVCS slit can be blue--shifted because the comet does not 
completely fill the slit, so the signal is coming primarily from a region close to the edge of the slit 
further from the Sun. Therefore, for the first comet exposure, the centroid shift is probably overestimated.

\begin{figure}[h]
\centering
    \includegraphics[angle=90,width=8.00cm]{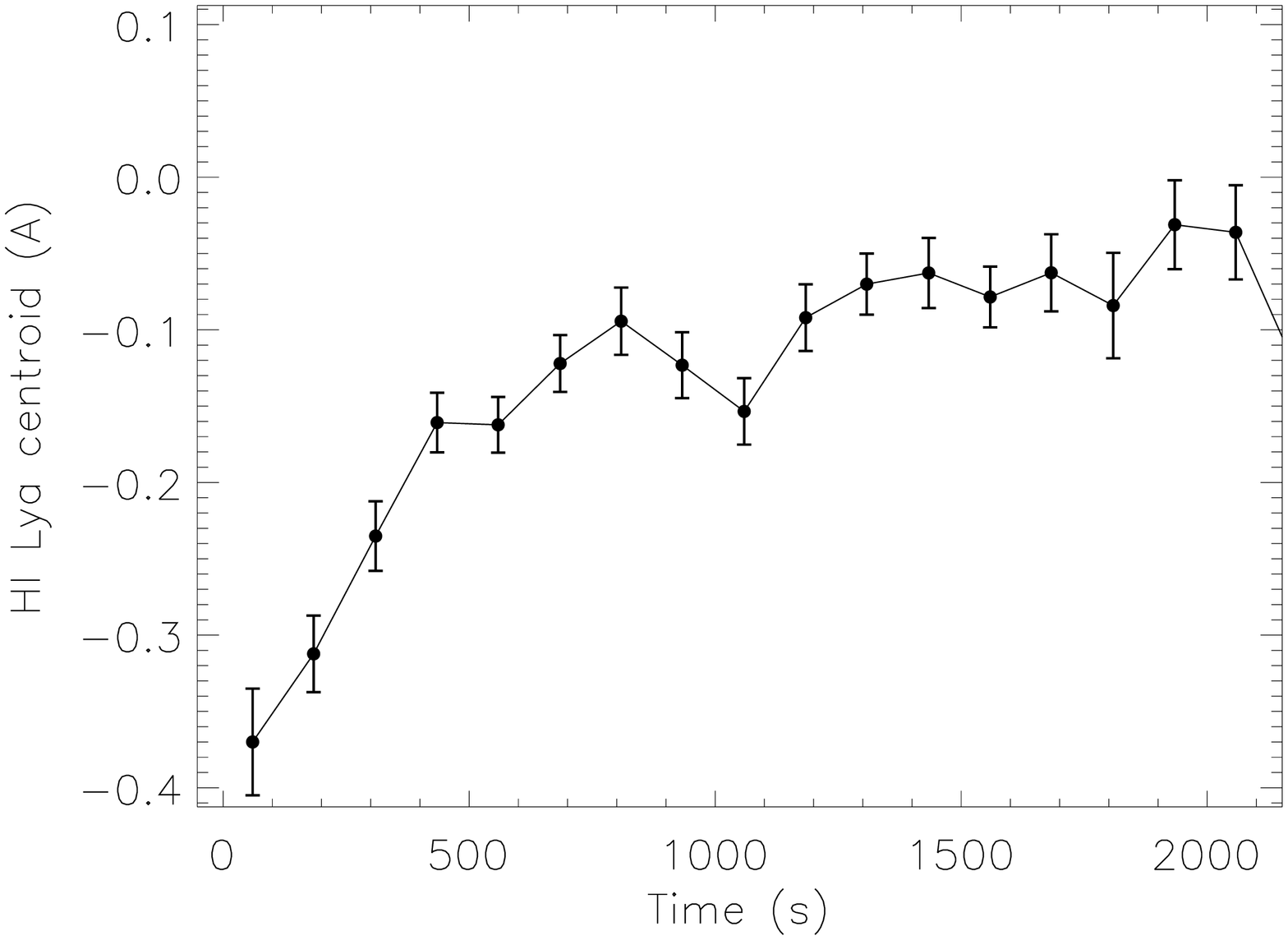}
    \includegraphics[angle=90,width=8.00cm]{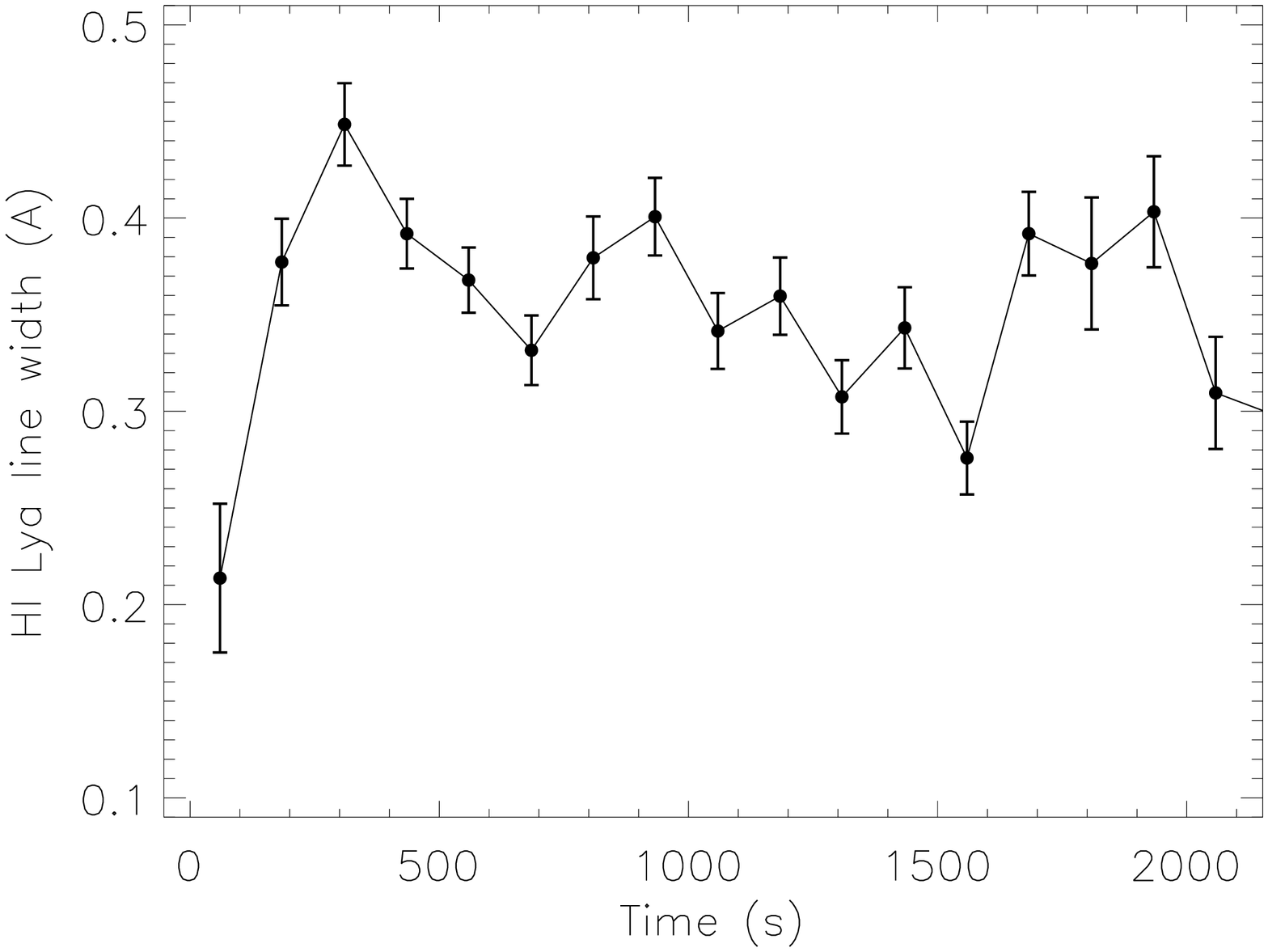}
  \caption{Observed \hi\, \lya\, centroid (upper panel) with respect to the background corona and 
  		\hi\, \lya\, line width (bottom panel) as a function of time integrated over a spatial region of $\pm$300" 
		around the comet axis from observations at $\rho_{obs}$=4.66~\rsun} 
  \label{fig:centroidwidth}
\end{figure}

We also compare the line width of \hi\, \lya\, line with the values from background exposures.
The comet velocity distribution is always comparable with the background.
No trend is evident as a function of time, except for the first exposure at each height.
The clearest difference from background is observed at 4.66~\rsun\, (see right panel of Figure~6), 
here we point out that the narrow line width in the first exposure is due either to partial filling of the UVCS aperture or 
to the pre--charge transfer component of \hi\, near the comet nucleus.
The line width gives information about the \hi\, \lya\, photon production mechanism.  
On the one hand, if the signal comes from solar radiation scattered by  \hi\, atoms created by photo--dissociation 
of water, we expect a narrow line profile because of the low speeds of those atoms.  
On the other hand, line widths close to that observed in the ambient corona are expected for photons 
scattered by \hi\, atoms after charge transfer with coronal protons.  

We reconstruct 2--dimensional comet images from UVCS spectral data using the known comet velocity 
in the plane of the sky, $V_{pos}$, based on LASCO observations and ephemeris computation. 
At each of the 4 observed heights the comet is detected as it crosses the UVCS slit at a fixed 
distance from the Sun.  For each exposure the intensity of the \hi\, \lya\, line at each position 
along the slit is measured.  Then the exposures are shifted in the radial direction by 
$r_{exp} = V_{pos}~(t_{exp} - t_{enter})$ where $t_{exp}$ is the time of the beginning of the 
exposure and $t_{enter}$ is the time of the first comet observation at each height.
Because the comet motion is perpendicular to the slit we do not need to shift the exposures in 
the direction parallel to the slit length.

The images of the UV comet emission from neutral hydrogen are shown in Figure~7.
All the images are 600" wide centered on the radial axis at polar angle of 236$^{\circ}$ counterclockwise from north pole, 
and the extent in the direction perpendicular to the comet path for the different heights depends on the comet velocity 
and the number of available exposures.
The \hi\, \lya\, tail due to interaction of outgassed neutral hydrogen with the coronal protons, widens with time because of 
the random thermal motions of the second generation \hi\, atoms.
  
\begin{figure}[h]
    \includegraphics[height=4.50cm]{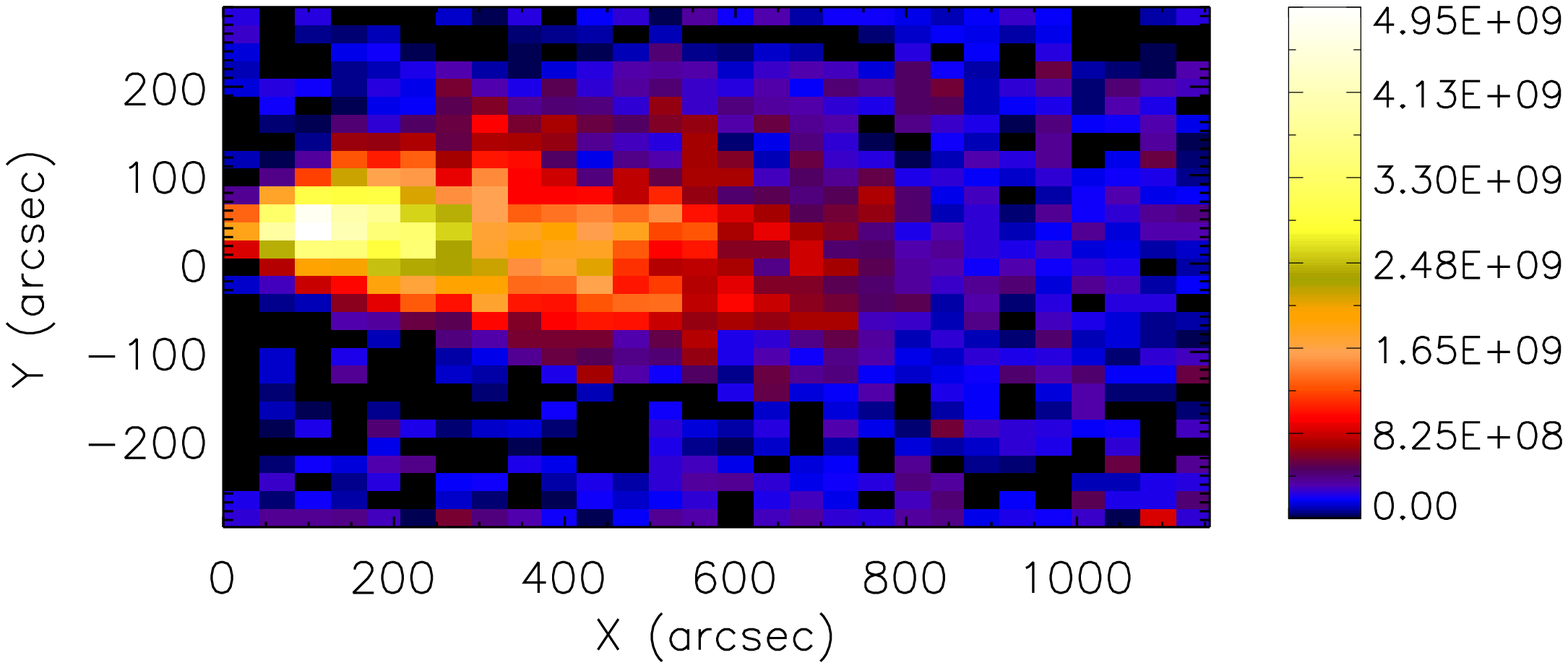}
    \includegraphics[height=4.50cm]{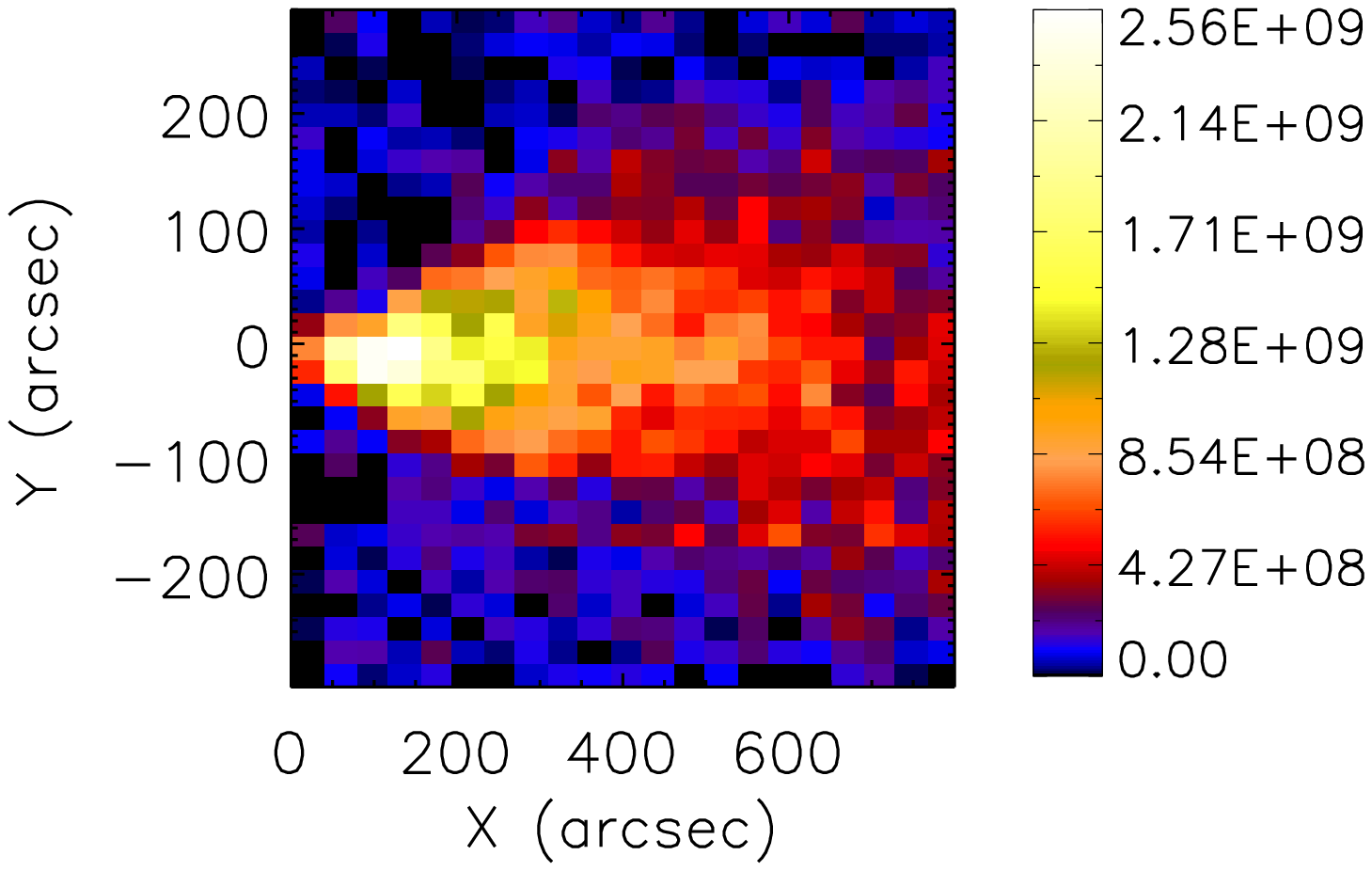}
    \includegraphics[height=4.50cm]{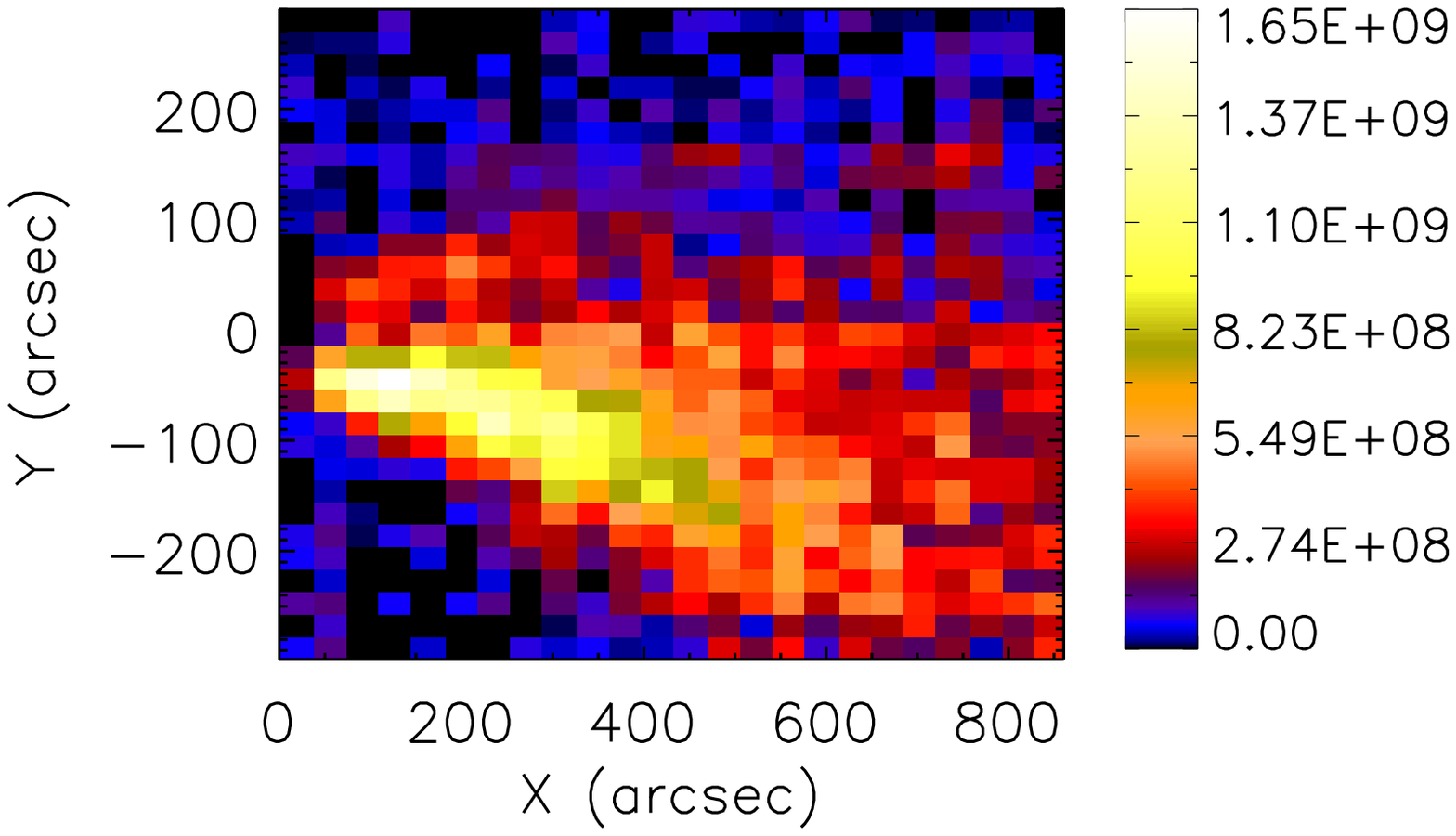}
    \includegraphics[height=4.50cm]{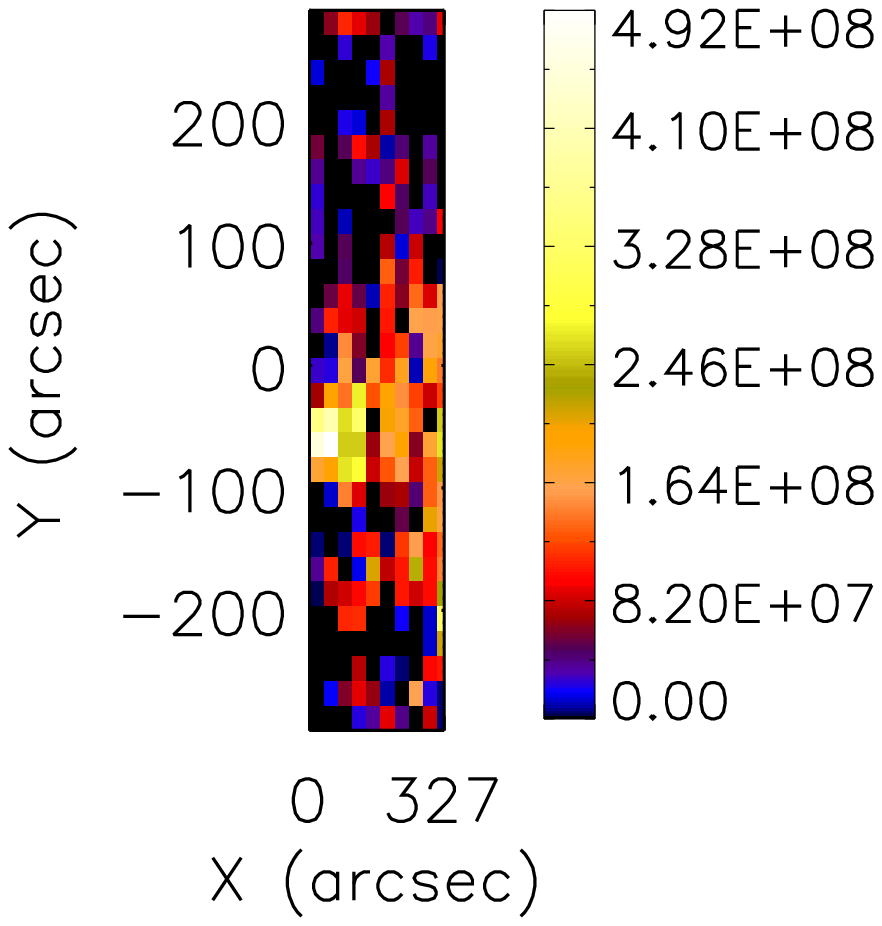}
  \caption{Comet \hi\, \lya\, intensity images from UVCS observations at 
    	      $\rho_{obs}$=4.66~\rsun, 
               $\rho_{obs}$=5.72~\rsun, 
               $\rho_{obs}$=6.84~\rsun\,  and 
               $\rho_{obs}$=7.40~\rsun\, from top to bottom.}
\end{figure}

The reconstructed 2--dimensional images obtained by the UVCS spectrometer can be compared to the images 
obtained in visible light by LASCO. 
We show the same portion of the sky observed from UVCS and LASCO C2 in Figures~8.  
We can see that from Sep 18 21:27 to 22:21 UT, UVCS clearly observes two comet tails at 6.84~\rsun, 
whereas a single, very narrow tail is imaged at 6.90~\rsun\, by LASCO C2 in a 25 seconds exposure at 21:30 UT. 
The LASCO tail is presumably the dust tail, which is much brighter than the ion tail in other sungrazing comets \citep{cia10}.  
The northern part of the UVCS image seems to correspond to the LASCO tail, whereas the southern part deviates 
from the comet trajectory.

\begin{figure}[h]
  \centerline{\epsfig{file=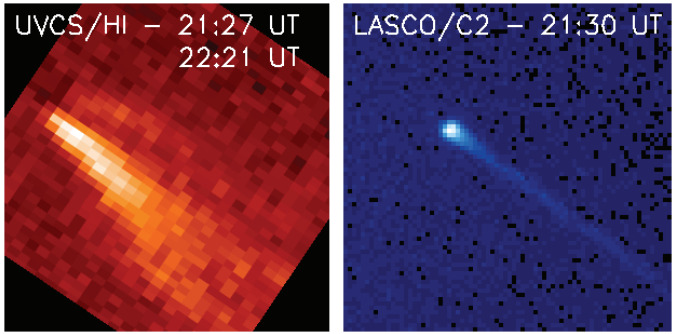,width=4.50cm,angle=270}}
  \caption{Comet image in \hi\, \lya\, from UVCS (left) and in white light from LASCO/C2 (right) 
  for the same region of sky at approximately the same time. 
  The size of the images is 855$\times$855".
  The image pixel size is 11.4" for LASCO/C2 telescope, while for UVCS it is 21~arcsec along the slit and 
  is $\sim$33" in the direction parallel to the comet path, the latter value is due to the comet movement 
  during the exposure time.}
\end{figure}

At the three lower heights with a suitable spatial binning the signal--to--noise ratio is adequate to study 
the structure of the \hi\, \lya\, line centroid along the slit by fitting the spectral lines.  
Therefore, we reconstruct the 2D Doppler images of the comet, that is the line centroid deviation 
from the background values.
As displayed in Figure~9, at all the heights, the upper part of the comet (closer to the equator) 
is blue--shifted with respect the background and the lower part shows a smaller, but clear red--shift, 
which is mainly evident at 6.84~\rsun .
At all heights, the blue--shifts reach values $>$100~\kms, whereas the red--shifts reach about 80~\kms.
We discard the hypothesis of an instrumental effect because the profiles along the slit of the line 
centroid of the background 
exposures do not show any significant trend, and only the first exposures are expected blue--shifted 
for instrumental reasons, as shown in the left panel of Figure~6.

\begin{figure}[h]
    \includegraphics[height=4.5cm]{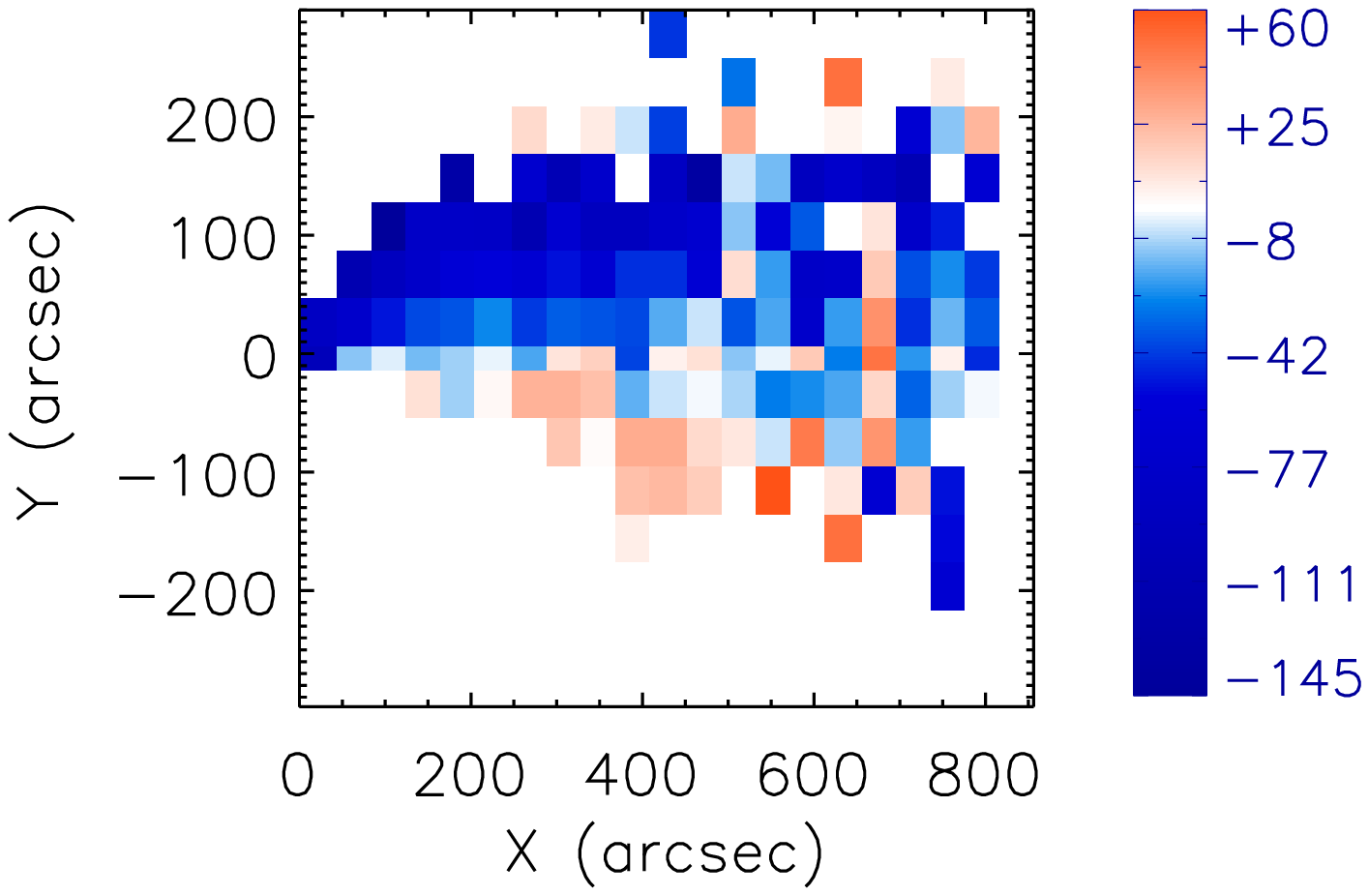}
    \includegraphics[height=4.5cm]{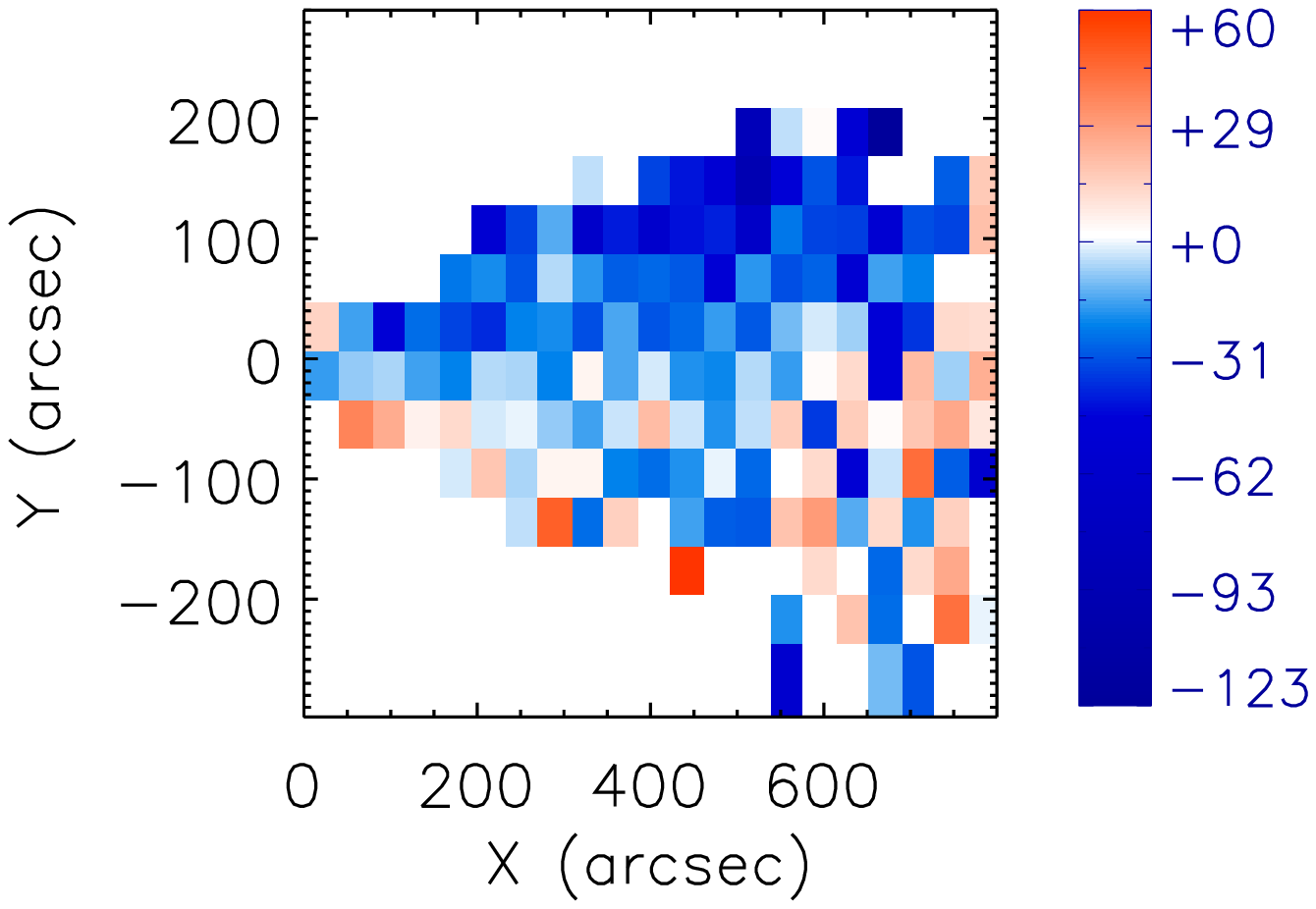}
    \includegraphics[height=4.5cm]{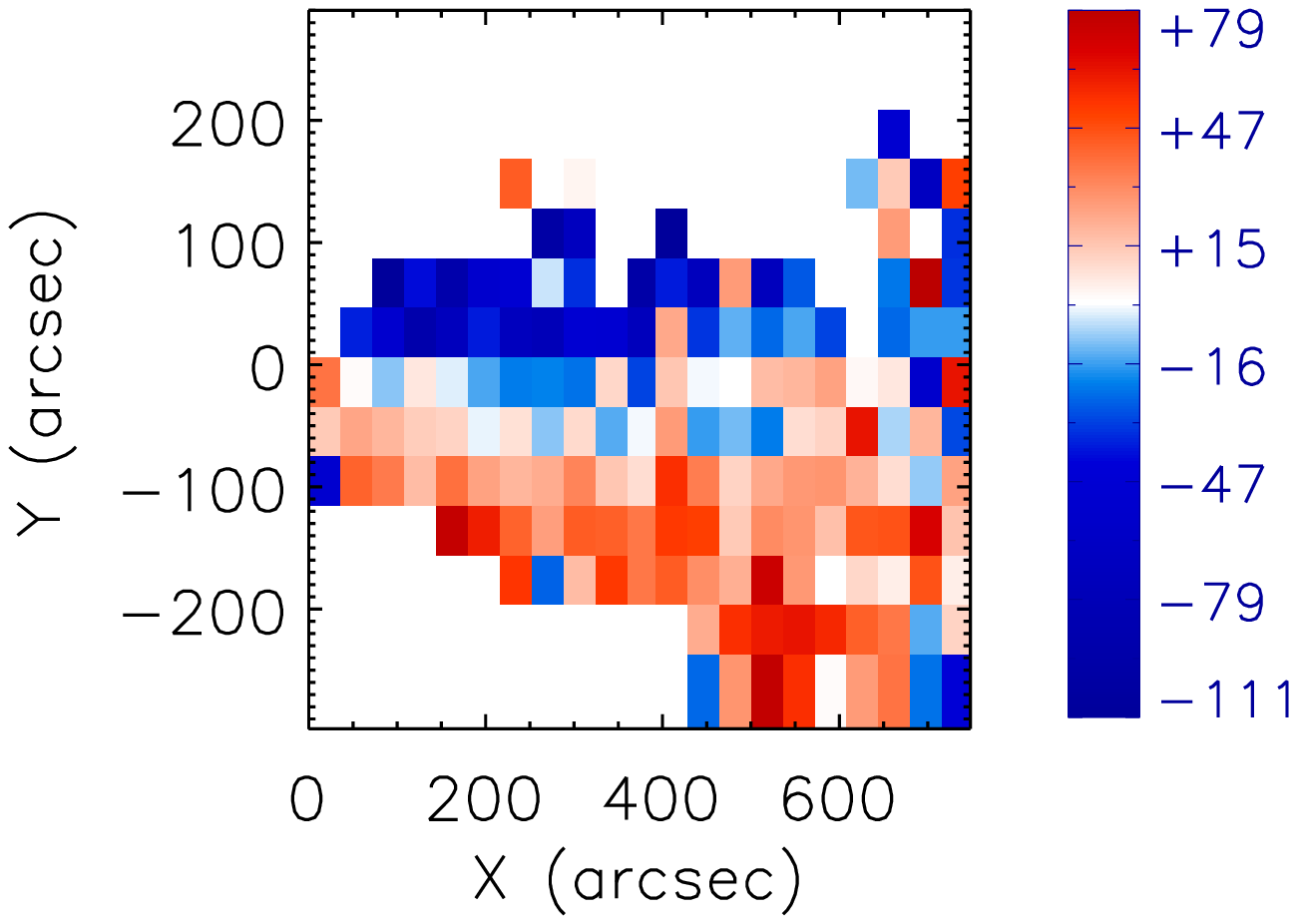}
  \caption{Doppler speed images of the \hi\, \lya\, comet emission from UVCS observations at 
  	      $\rho_{obs}$=4.66~\rsun\, (upper), 
               $\rho_{obs}$=5.72~\rsun\, (center) and $\rho_{obs}$=6.84~\rsun\, (bottom). 
               The color bars show the Doppler speed in \kms.}
\end{figure}

\section{Analysis}

Previous studies of sungrazers observed by UVCS \citep{uzz01, bem05} derived the parameters
of the comet and of the solar wind based on a simple semi--analytic model which assumes that 
the outgassed \hi\, atoms travel with the nucleus and that a single timescale describes the
ionization and charge transfer processes. In this work, we compare the observations with a 
more detailed model based on Monte Carlo simulations.  Before describing the simulations,
we summarize the physical processes involved, the rates we assume, and the likely ranges of 
the parameters of the comet and coronal plasma.

\subsection{Physical Processes and Rates}

As a comet approaches the Sun, water is very quickly
photodissociated into hydrogen and oxygen atoms through the reactions:
\begin{equation}
\water + h\nu\rightarrow \oh + H
\end{equation}
\begin{equation}
\oh + h\nu\rightarrow \oo + H.
\end{equation}
creating a first generation of neutral H atoms with speeds of
8 (from OH) to 20 (from $\water$) $\rm km~s^{-1}$  \citep{com88,mak01B} with respect to the nucleus.
If the density is high enough, collisions between hydrogen atoms, molecules and heavier atoms will slow the
H atoms to a few $\rm km~s^{-1}$.
These H atoms can resonantly scatter chromospheric \hi\, \lya\, photons,
but because they are more or less moving at the speed of the comet, the
scattering is subject to strong Doppler dimming (Swings effect) at the comet radial speeds
above the half width of the chromospheric emission profile, or about 100 \kms\, \citep{lem02}.
\citet{noc99} give the cross section and angular dependence of this process.
As discussed in greater detail below, the population of neutrals formed by photodissociation has
a modest optical depth near the comet, which will reduce the number of scattered photons by 5\% - 50\% for the
outgassing rates and solar wind densities considered here.

The first generation H atoms outgassed from cometary nucleus are subject to
collisional ionization by electron impact and photoionization, which reduces the neutral atom population,
but most will undergo charge transfer with coronal protons and generate a second generation population.
At typical coronal electron temperatures ($T_e\approx10^6$ K) and densities
($n_e\approx10^3-10^5$ $\rm cm^{-3}$), the collisional ionization rate  is
\begin{equation}
\tau_{coll}^{-1}=\gamma~n_{e} \rm~s^{-1}
\end{equation}
where $\gamma\approx 2.78~\times~10^{-8}\rm~cm^3~s^{-1}$ \citep{scho91},
and the photoionization rate is
\begin{equation}
\tau_{ph}^{-1} \approx 2 \times 8.3 \times 10^{-5} ~\biggl(\frac{6.8}{r}\biggr)^{2} \rm ~s^{-1}
\end{equation}
as calculated by \citet{ray98B} for a comet observation at solar minimum, where $r$ is the heliocentric
distance in \rsun.   We have increased the photoionization rate by a factor of two to account for the increased 
UV emission near solar maximum.
The charge transfer process between the first generation neutral H atoms and the solar wind protons produces a
population of neutral hydrogen with nearly the same velocity distribution, kinetic temperature and bulk speed
as the solar wind protons with a rate

\begin{equation}
\tau_{ex}^{-1}=(\sigma_{ex}~n_{e}~v_{eff}) \rm ~s^{-1}
\end{equation}

\noindent
where $\sigma_{ex}~=~1.3~\times~10^{-15} cm^{2}$ \citep{schu08}, is the charge transfer cross section and
$v_{eff}$, the effective velocity for charge transfer, is the sum (in quadrature) of the proton thermal
speed, $V_{th}$ , and the comet speed relative to the solar wind, $V_{rel}~=~V_r-V_w$, where $V_w$ is the
solar wind speed.  At the observed heights, charge transfer is the most rapid process.
The post--charge transfer neutrals also scatter chromospheric \hi\, \lya\, and the similarities between the cometary
and coronal \hi\, \lya\, profiles indicate that these second generation H atoms are the main source of
the \hi\, \lya\, comet emission.
Collisional excitation of \hi\, \lya\, is generally less important than photoexcitation at the densities and
heliocentric distances considered here, even when Doppler dimming is strong.
From an observational point of view, the collisional excitation can be discarded by considering that the
\hi\, \lyb\, emission from this comet is negligible, as was found in previously observed sungrazers
\citep{ray98B,uzz01}.  The ratio of Ly$\beta$ to Ly$\alpha$ is about
0.001 for radiative excitation and 0.16 for collisional excitation, so the non--detection Ly$\beta$ implies that
the collisional contribution to Ly$\alpha$ is small.

These processes, along with the neutral H outgassing rate,  $\dot{N}_H=2\dot{N}_{\small \water}$, and the coronal
parameters, such as the electron density, $n_e$, and the solar wind outflow speed, $V_{w}$, determine the
cometary \hi\, \lya\, light curve.  The temperature of the corona also enters, in that the extent
of the \hi\, \lya\,  emission along the UVCS slit depends the speed of the second generation neutrals,
and therefore on the temperature of the coronal protons from which
they form.  The coronal proton kinetic temperature in principle can be determined from observations of the
background corona before the comet crosses the slit, but the values obtained are line--of--sight averages that
may differ from the values at the actual position of the comet.
In addition, the measured coronal spectra include narrow background and foreground interplanetary \hi\, \lya\, and
stray light from the solar disk not fully suppressed by the telescope occulters \citep{Cra10}.
For this reason, we simulated three different temperatures from $0.8$ to $1.5$ $\times 10^6$ K.
Based on other sungrazing comets we expect $\dot{N}_H$ $\sim$ $10^{28}$ to $10^{29}$ per second.
The expected electron densities are in the range from current sheet to coronal hole values at solar maximum from \citet{guh06}.
We expect wind speeds of 50 to 250 $\rm km~s^{-1}$ from the UVCS observations of \citet{str02} and the LASCO
observations of \citet{she97}.
Higher wind velocities would imply such strong Doppler dimming that the resonantly scattered \lya\,
would be faint, but the \lya\, to \lyb\, ratio shows that this is not the case.

We neglect solar radiation pressure and gravity effects in the computation 
because they are very small for sungrazing comets with respect to the above processes. 
We also neglect the effect of the cometary material on the dynamics of the solar wind and 
the interaction with the magnetic field.  
We address the latter effect in the Discussion section.

\subsubsection{Optical Depth}

One important physical correction to the earlier analyses of UV observations
of sungrazing comets \citep{ray98B,uzz01,bem05} is the effect of finite optical depth of
the \hi\, \lya\, line.
For larger comets at greater distances from the Sun, detailed radiative transfer calculations
have been performed by \citet{com88}, and the optical depth has been inferred
from the \hi\, \lya\, intensity profile \citep[e.g.,][]{ray02}
A cloud of hydrogen expanding at a speed V has a density $\dot{N}_H/4 \pi r^2 V$, and the
\hi\, \lya\, scattering cross section is inversely proportional to the velocity width.
Therefore, the opacity is proportional to $V^{-2}$ and optical depth effects are far more important
for the first generation atoms ($V \sim 10~\rm km~s^{-1}$) than for the second
($V > 100~\rm km~s^{-1}$).

To model the effects of opacity in the \hi\, \lya\, line, we assumed a spherical cloud with $\dot{N}_H$
particles per second injected at the center with an expansion velocity V and a radius $r_c$  at which
the neutrals are destroyed by ionization or removed from the slowly expanding population by charge transfer.
The radius of that cloud is $r_c = V \times \tau_{d}$,
where $\tau_{d}=(1/\tau_{ph} + 1/ \tau_{coll} + 1/\tau_{ct}  )^{-1}$ is the lifetime of neutral atoms.
The density is
\begin{equation}
n~=~\frac{\dot{N}_H}{4 \pi r^{2} V}~ e^{-r/r_c}
\end{equation}

The \hi\, \lya\, scattering cross section averaged over the line profile is
\begin{equation}
\sigma~=~0.026 f \lambda / \Delta V~\sim~1.3 \times 10^{-7} / \Delta V~~\rm cm^2
\end{equation}
where $f$ is the oscillator strength,  $\lambda$ is the wavelength of the transition and
$\Delta V$ is the FHWM of \hi\, \lya\, line.

For a grid of values of the coronal density, wind speed,
and $\dot{N}_H$, we compute the suppression factor
\begin{equation}
S_{fact} =\frac{\int n e^{-\tau}}{\int n}
\end{equation}
for density, $n$, and optical depth, $\tau = \sigma \times \int n$, averaged over the cloud.

As mentioned above, the optical depth is important only for the first generation neutrals, and the expansion
speed, hereafter called attenuation speed, $V_{att}$, is a critical parameter.
Hydrogen atoms are formed at speeds of 8 to 20 $\rm km~s^{-1}$ by photo--dissociation
of $\water$ and OH.  If the dissociation occurs where the density is low, they expand with that range of speeds.
However, much of the photodissociation can occur where the density is high enough that the gas is collisional,
especially at the high temperatures expected close to the Sun.
For the observation at 4.66~\rsun\, the suppression factor of \hi\, \lya\, for the pre--charge transfer component
has been computed for grids of models with $V_{att}$ = 16.0, 5.0, 3.5 $\rm km~s^{-1}$,
in the $\dot{N}_H$ range from $10^{27}$ to $10^{30}~\rm s^{-1}$,
$V_w$ from 50 to 250 $\rm km~s^{-1}$ and
coronal density from $1 \times 10^{3}$ to $5 \times 10^{4}~\rm cm^{-3}$.
For example, with
$\dot{N}_H=2 \times 10^{28}~\rm s^{-1}$,
$V_{w}=100~\rm km~s^{-1}$ and
$n_e = 1.75 \times 10^{3}~\rm cm^{-3}$
the suppression factors are 0.71, 0.25 and 0.15 for 
$V_{att}$ = 16, 5 and 3.5 $\rm km~s^{-1}$, respectively.
This suppression factor is applied to scattering from all first generation atoms, while second generation
atoms are assumed to be in the optically thin regime in the Monte Carlo simulations.

\section{Monte Carlo Simulation}

In order to better understand the relationships between the observed \hi\, \lya\, spectral emission and
the physical parameters of the coronal plasma encountered by the comet we developed a simulation code 
based on the Monte Carlo technique.
The purpose is to reproduce the observed sungrazer  \hi\, \lya\, spectra, and therefore the reconstructed 
comet light curves and images, as functions of coronal and cometary parameters.

At each observed height we run a large number of simulations for a grid of coronal and cometary input parameters.  
The coronal parameters are the electron density, $n_{e}$, the wind 
velocity, $V_{w}$ and the proton kinetic temperature, $T_{k}$.
We choose different values, $n_{e,0}$ and $V_{w,0}$, at the comet observed 
radial distance, $r_{0}$, then we define the electron density 
radial profile as given by \citet{guh06} and the wind velocity radial 
profile from mass flux conservation.  The kinetic temperature is
assumed constant in the radial range of interest.
The only cometary grid parameter is the neutral hydrogen outgassing
rate at the comet observed distance, $\dot{N}_{0}$, 
which is assumed to decrease inversely proportional to square distance:  
$\dot{N}_H(r)=\dot{N}_{0} (\frac{r_0}{r})^2$.

A simulation run begins with a generation of a sample of neutral hydrogen atoms
at a heliocentric distance far from the UVCS field of view, 
moving with the bulk velocity of the comet 
plus a spherically symmetric outgassing velocity distributed in the 
range 8 to 20~\kms\, \citep{com88,mak01A}.
We start the simulations far enough from the UVCS slit that
less than 1\% of the number of particles generated at the first time step reaches the UVCS field of view.
The comet position and its kinematic parameters evolve as functions of time 
following the trajectory computed from orbital parameters (see Tables 1 and 2).
At each simulation time step, $\delta t = 5 s$,  a number of new particles 
(H atoms) proportional to $\dot{N}_H(r)$ is generated at the 
new position, $r$, and added to the previous particles. 
Then the simulated atoms can be subject to two different processes; ionization 
or charge transfer, as discussed in Section~3.1.
Based on the position of each particle and the coronal electron density, we compute
the collisional, $\tau_{coll}^{-1}$, and photoionization $\tau_{ph}^{-1}$, rates 
(Equations~3 and 4), and the ionization probability for each particle in 
the $\delta t$ time interval is $P_{ion}= 1- exp[-(\delta t/\tau_{ion})]$, 
where $\tau_{ion}^{-1}=\tau_{coll}^{-1}+\tau_{ph}^{-1}$.  With the Monte Carlo 
method we statistically remove the ionized atoms, reducing their number.  Similarly, 
the charge transfer probability is driven by the rate given in Equation~5, where we 
need to account for the comet, wind and coronal thermal velocities at the position 
of each particle. 
All the particles that undergo charge transfer get new velocities given by 
the composition of the radial wind velocity and the random thermal speed.
At each time step we update the position $\vec{P}(x,y,z)$ and velocity $\vec{V}(v_x,v_y,v_z)$ 
of all particles in the three--dimensional space, 
where $x$ and $y$ defines the plane of the sky and $z$ the line--of--sight direction.

\subsection{Computation of \hi\, \lya\, Cometary Emission}

The only relevant mechanism to produce \hi\, \lya\, emission is resonant scattering 
of the chromospheric radiation.  This process is characterized by the so--called 
{\it{g-factor}} \citep[e.g.,][]{oxenius65,mak01A}, the number of photons per second scattered 
by each atom, which for \lya\ resonance of neutral hydrogen at 1 AU can be expressed as

\begin{equation}
{g}_{AU} = \frac{\pi e^2}{m_e c}f_{\alpha} \frac{\lambda_0^2}{c}~F_{\odot}
\end{equation}

\noindent
where $f_{\alpha}$ is the  \hi\ \lya\, oscillator strength,
$\lambda_0$ is the reference wavelength in $nm$, $F_{\odot}$ is exciting 
specific solar flux in $ph~cm^{-2}~s^{-1}~nm^{-1}$,
$e$ and $m_e$ are the electron charge and mass.
We scale the previous equation to the heliocentric distance, $r$ measured in \rsun\, units, 
of each simulated scattering particle by the dilution factor,
$\Omega_{r}/\Omega_{AU}$, 
that is the ratio between the solid angles subtended by the source of 
resonant radiation at 1 AU and at the distance $r$. 
Following \cite{noci87}, we compute the solid angles as $\Omega=2\pi (1-\sqrt{1-\frac{1}{r^2}})$,
and we take into account the angular dependence of the \hi\ \lya\ scattering through the function
$p(\phi)=\frac{11+3\cos^{2}\phi}{12}$, 
where $\phi$ is the scattering angle between the Sun--particle vector and the line--of--sight.
Therefore, for each simulated particle we compute the number of scattered photons with:
\begin{equation}
{g} = \frac{\Omega_{r}}{\Omega_{AU}}~
\frac{11+3\cos^{2}\phi}{12}~
\frac{\pi e^2}{m_e c}f_{\alpha} \frac{\lambda_0^2}{c}~F_{\odot}
\end{equation}
where the \hi\ \lya\ exciting specific flux, $F_{\odot}$, measured by SUMER/SOHO \citep{lemaire2002},
is selected as the value corresponding to the radial velocity, $v_r$, of each scattering particle.
Finally, for each scattering particle we determine the emitted wavelength, $\lambda$, from the line--of--sight velocity, $v_z$, 
of the particle itself as $\lambda=\lambda_0(1-\frac{v_z}{c})$, where $\lambda_0$ is the rest wavelength 
of the \hi\ \lya\ transition.

From the number of photons scattered per second by each simulated particle 
we compute the emitted spectra by combining, along the line--of--sight (los), 
all the photons from all the atoms in each simulated spatial element defined as the UVCS spatial resolution. 
The emission per unit of wavelength is obtained by collecting all the photons emitted in each spectral window 
with resolution $\Delta\lambda$=0.09\AA. 
In this way, for each spatial element in the plane of the sky ($\Delta x \Delta y$), 
we determine the \hi\, \lya\, specific intensity, 
$I(\lambda)$ in $ph~cm^{-2}~s^{-1}~sr^{-1}~A^{-1}$ as
\begin{equation}
I(\lambda)=\frac{1}{4\pi}\frac{1}{\Delta x ~\Delta y}\frac{1}{\Delta\lambda}\sum_{los}g
\end{equation}
where the factor 1/$4\pi$ normalizes the emission to unit of solid angle and $1/\Delta\lambda$ to unit of wavelength.
Because the simulated time steps are shorter than the UVCS exposure time, we average the intensity 
from all the time steps over the UVCS integration time of a single exposure.  Finally, the simulated 
spectra are multiplied by the ratio between the assumed outgassing rate,  $\dot{N}_{0}$, and the 
number of simulated outgassed particles per second.  In this way we obtain intensity spectra 
directly comparable to the radiometric calibrated spectra from UVCS observations. 

\subsection{Model Grid of Parameters}

Table~3 summarizes the grid of the free parameters used to simulate the comet observation at the lowest height, 
$\rho_{obs}$=4.66~\rsun, which corresponds to an actual heliocentric distance r=5.99~\rsun. 
The outgassing rate, at the comet observed distance, assumes values from $10^{27}$ to $10^{30}$ $s^{-1}$ 
as expected from previous UVCS sungrazing comet observations \citep{ray98B,uzz01,bem05,cia10}
We ran simulations for wind velocities from 50 to 200 \kms, with 25 \kms\ steps and also with $V_{w,0}$=0 \kms\, 
as a control value which is not physically significant, and with $V_{w,0}$=250 \kms , which 
ultimately develops a very faint signal because of Doppler dimming. 
The electron density can assume ten different values from 0.05 to 1.92 times the 
radial profile determined by \citet{guh06} 
for coronal current sheets, and three different values  
($8 \times 10^{5}$, $1.1 \times 10^{6}$ and $1.5 \times 10^{6}$ K) 
for coronal kinetic temperature are modeled.
We ran a model neglecting the optical depth and three models with different attenuation speeds 
(3.5, 5.0 and 16.0~\kms, see section 3.1.1), giving different suppression factors for the emission from 
pre--charge transfer atoms.

\begin{table} [h]
\caption {Simulation grid free parameters at $\rho_{obs}$=4.66 \rsun}
\centerline{
\begin{tabular}{|l|l|l|c|}
\hline
Quantity	& Description	& Range	& Steps  \\
\hline
$\dot{N}_{0}$ & Outgassing rate	& $10^{27}$ to $10^{30}$ atoms $\rm s^{-1}$ & 61 \\
$V_{w,0}$  & Wind velocity		& 0 to 250 \kms                                & 9   \\
$n_{e,0}$	   & Electron density 	& $1.1\times 10^{3}$ to $4.1\times 10^{4}$  $\rm cm^{-3}$  &  10 \\
$T_{k}$	   & Proton temperature	& $0.8\times 10^{6}$ to $1.5\times 10^{6}$ $K$	 & 3  \\
$V_{att}$	   & Attenuation speed  & 0 to 16 $\rm km~s^{-1}$	& 4     \\
\hline
\end{tabular}
}
\end{table}

\subsection{Simulation--Observation Comparison}

For all the possible combinations of these free parameters we simulate the \hi\ \lya\ 
spectra emitted by the comet crossing the UVCS slit field of view. 
By integrating the spectra over the wavelength and over the spatial direction along the slit we obtain 
the total intensity as a function of time as the comet crosses the UVCS slit, that is the simulated light curves 
comparable with the  those observed (see Figure~5).  
Then we compare the observations with each simulation by computing the $\chi_{\nu}^2$ defined as
$$
\chi^{2}_{\nu}~=~\frac{1}{\nu} ~\sum_{N}\frac{\left(I_{obs} - I_{sim}\right)^2}{\sigma_{obs}^2 + \sigma_{sim}^2}
$$
where $N$ is the number of data points, $\nu=N-n-1$ is the number of degrees of freedom with $n=5$ is the number 
of parameters of the model, $I_{obs}$ and $I_{sim}$ are the observed and simulated intensities and 
$\sigma_{obs}$ and $\sigma_{sim}$ their statistical standard deviations. 
In this way we obtain a multi--dimensional array $\chi^{2}_{\nu}(\dot{N}_{0}, V_{w,0}, n_{e,0}, T_{k}, V_{att})$, 
which quantitatively summarizes the comparison of all simulated models with observations. 
Then minimizing this parameter determines the set of coronal and cometary parameters which
best fit the observations.

First of all, we search for the best fit for the four different corrections of the opacity (see Table~3). 
We found that the models with the intermediate speed values, $V_{att}$=3.5 and 5.0 \kms, produce 
comparable results, both matching the observed data at 4.66 \rsun\, quite well (see blue and red models in Figure~10), 
whereas the models which neglect the opacity and those with higher speed show spikes of narrow, 
red--shifted emission for 1 or 2 exposures when the comet enters the slit which are not seen in the data 
(Figure~10).
Therefore, we choose the $V_{att}$ = 3.5 $\rm km~s^{-1}$ model, which suppresses spike 
for all but the most extreme choices of other parameters.
We point out that the approximation of two hydrogen atom populations, from water (20~\kms) 
and from OH (8~\kms) photodissociation, should work well for larger heliocentric distances 
($>$ 0.8~AU) \citep{com05},
whereas at small heliocentric distances the velocity distribution emerging from the collision zone is expected 
to pile--up at low velocities ($<$ 4~\kms) \citep[e.g.,][]{com88}.

\begin{figure*}[h]
  \centering
    \includegraphics[height=8.00cm,angle=90]{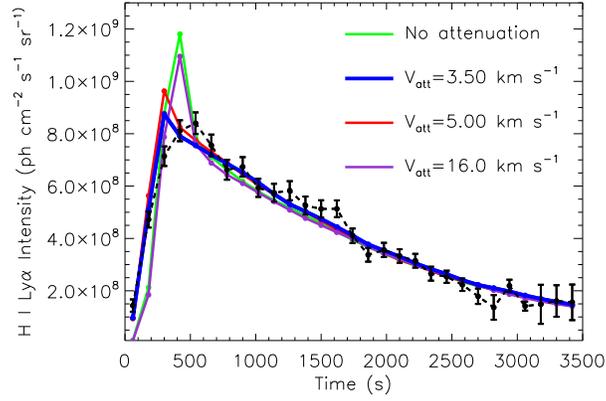}
  \caption{Observed and best--fit comet light curve at $\rho_{obs}$=4.66~\rsun\, 
  with different optical depth assumptions.
  Black dots with error bars are the observed \hi\ \lya\ intensity. 
  The green curve represents the best simulated light curve without correction
  for optical depth. The blue curve represents the correction with the attenuation 
  speed of 3.5 \kms, the red curve with 5.0 \kms\ and the violet one with 16 \kms.} 
\end{figure*}

The best fit between the observed and simulated comet light curves at $\rho_{obs}$=4.66~\rsun,
corresponding to the radial distance $r=5.99$ \rsun\, is obtained with the following coronal parameters; 
$V_{w,0}$=75 \kms, $n_{e,0}=1.23\times 10^{4} cm^{-3}$ and $T_{k}=8\times 10^{5}$ K
and with the comet outgassing rate $\dot{N}_{0}=1.12\times 10^{28}s^{-1}$.

To verify that the best fit corresponds to a minimum of the $\chi_{\nu}^2$ distribution,
we show the two--dimensional maps of $\chi_{\nu}^2$ for selected values of three parameters.
For example with fixed $T_{k}=8.0 \times 10^{5} \rm K$ and $V_{att}=3.5$ \kms\,
we show the $\chi^{2}_{\nu}(\dot{N}_{0}, V_{w,0})$ map with 
$n_{e,0}=1.23 \times 10^{4} \rm cm^{-3}$ (top panel of Figure~11), the
$\chi^{2}_{\nu}(\dot{N}_{0}, n_{e,0})$ map with $V_{w,0}=75$ \kms\ (middle panel) and the
$\chi^{2}_{\nu}(V_{w,0}, n_{e,0})$ map with $\dot{N}_{0}=1.12 \times 10^{28} \rm s^{-1}$  (bottom panel).
As shown in the maps, the $\chi_{\nu}^2$ distribution has a minimum 
around the values of the parameters which define the best fit.

\begin{figure*}[h]
  \centering
    \includegraphics[height=7.750cm,angle=90]{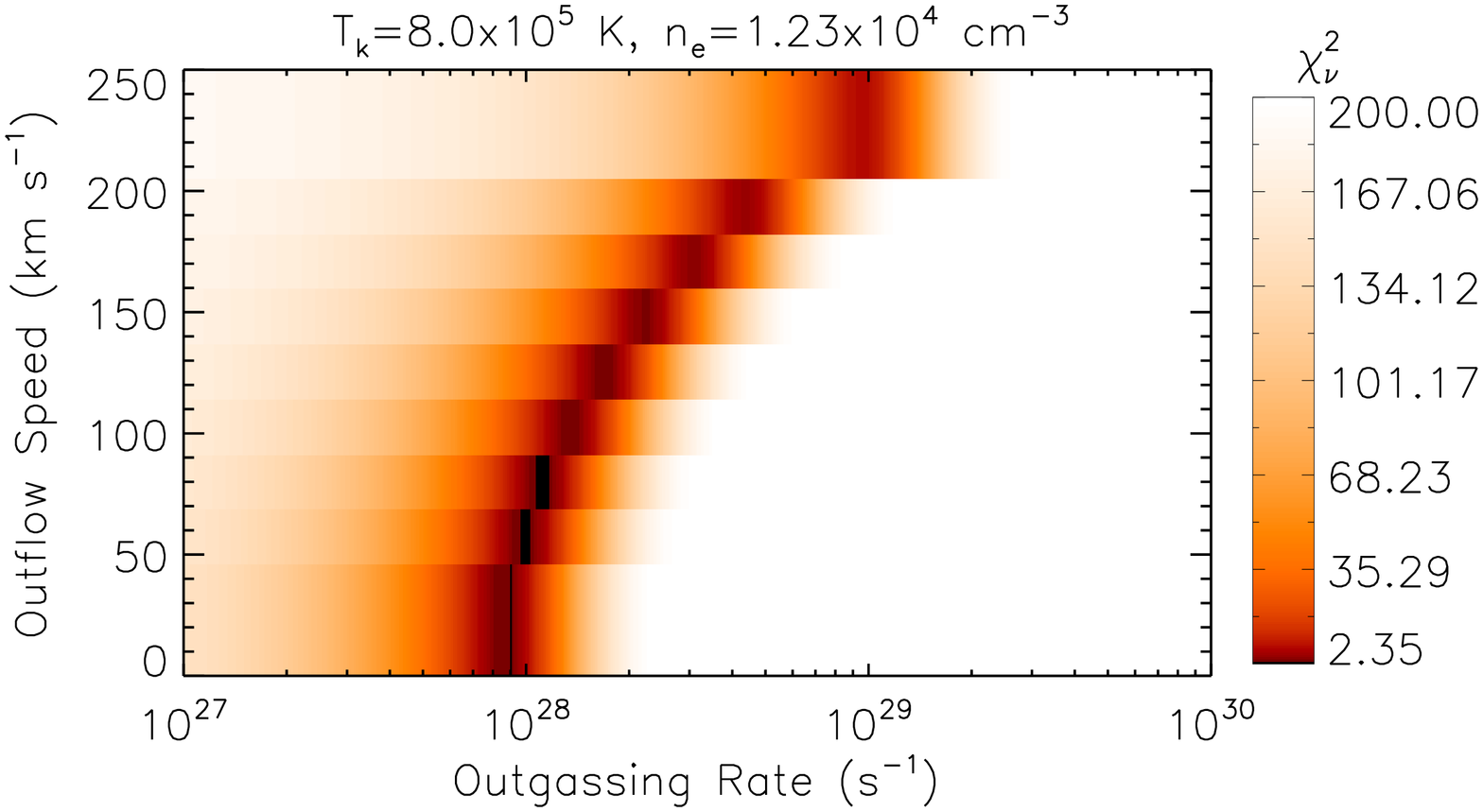}\\
    \includegraphics[height=7.750cm,angle=90]{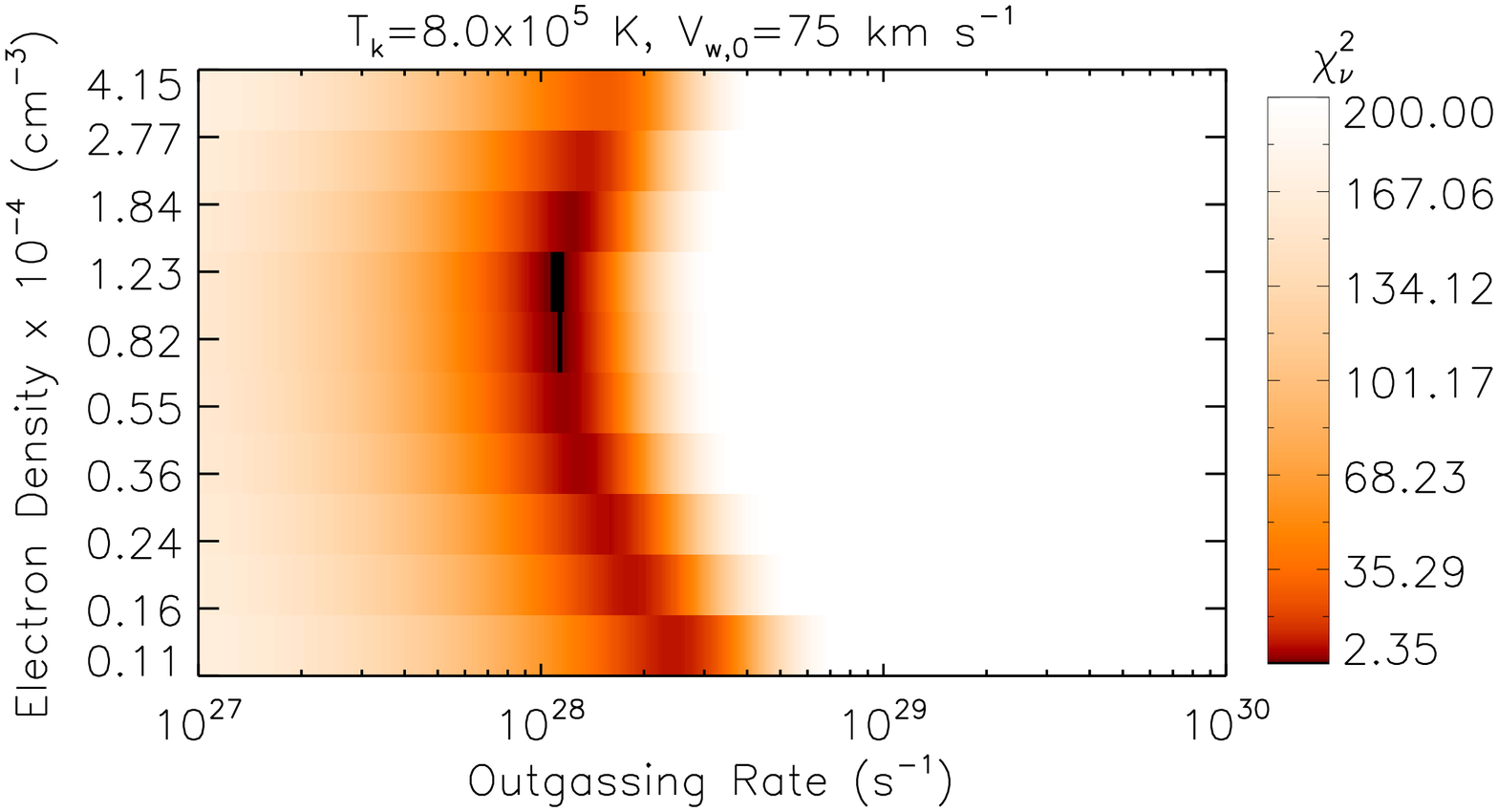}\\
    \includegraphics[height=7.750cm,angle=90]{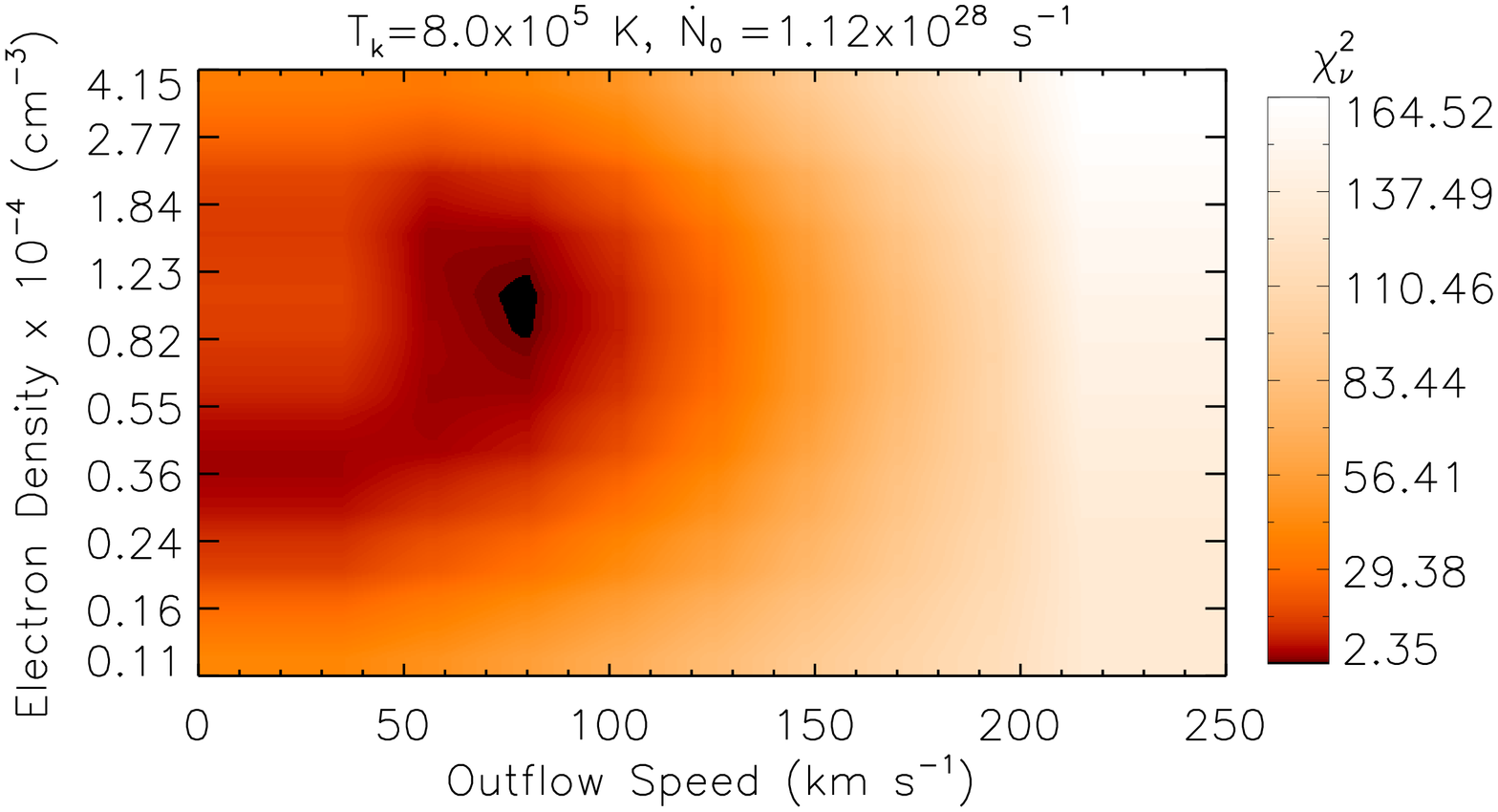} \\
  \caption{Example of $\chi_{\nu}^2$ maps. 
  Top: $\chi^{2}_{\nu}(\dot{N}_{0}, V_{w,0})$, 
  middle: $\chi^{2}_{\nu}(\dot{N}_{0}, n_{e,0})$, 
  bottom: $\chi^{2}_{\nu}(V_{w,0}, n_{e,0})$ } 
\end{figure*}

\begin{figure*}[h]
  \begin{tabular}{cc}
   \hspace{0.0cm} \includegraphics[height=4.55cm]{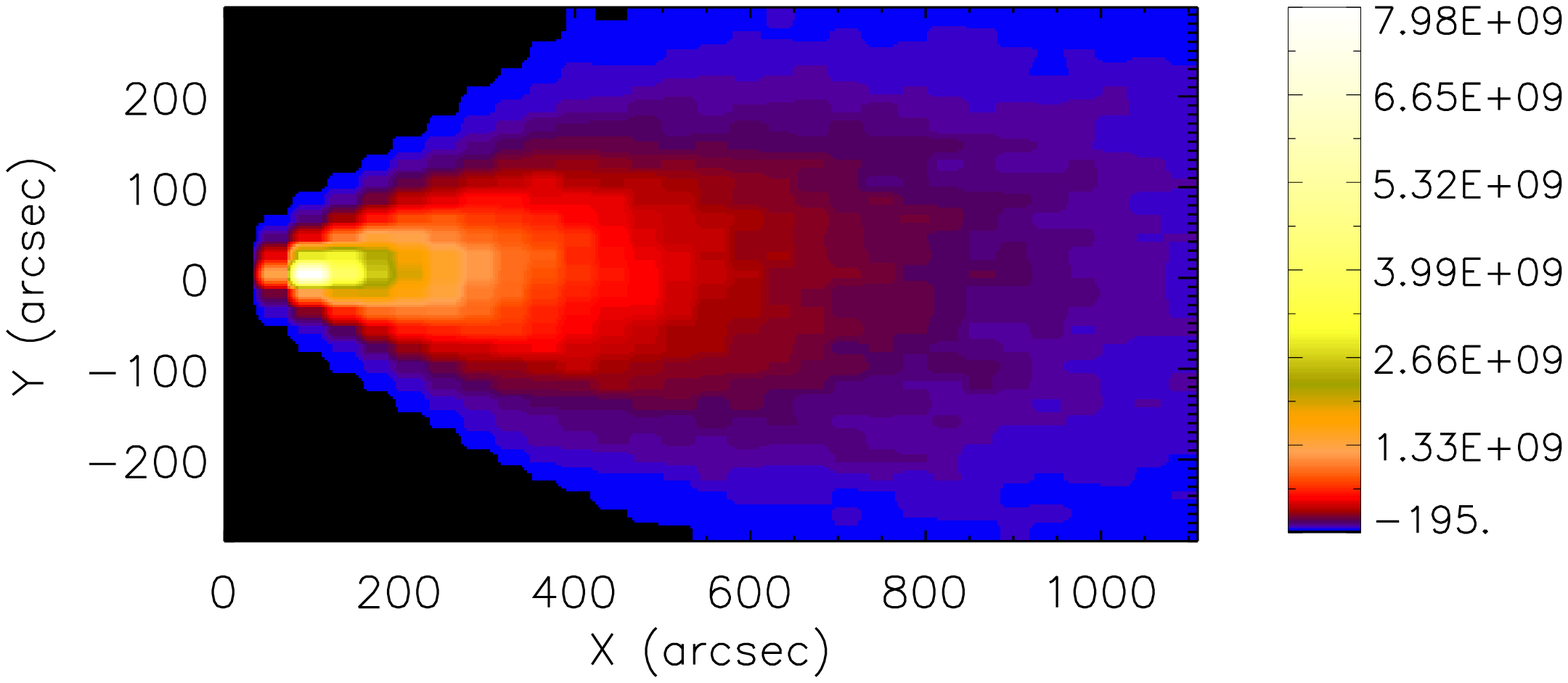} &
   \hspace{0.0cm} \includegraphics[height=4.55cm]{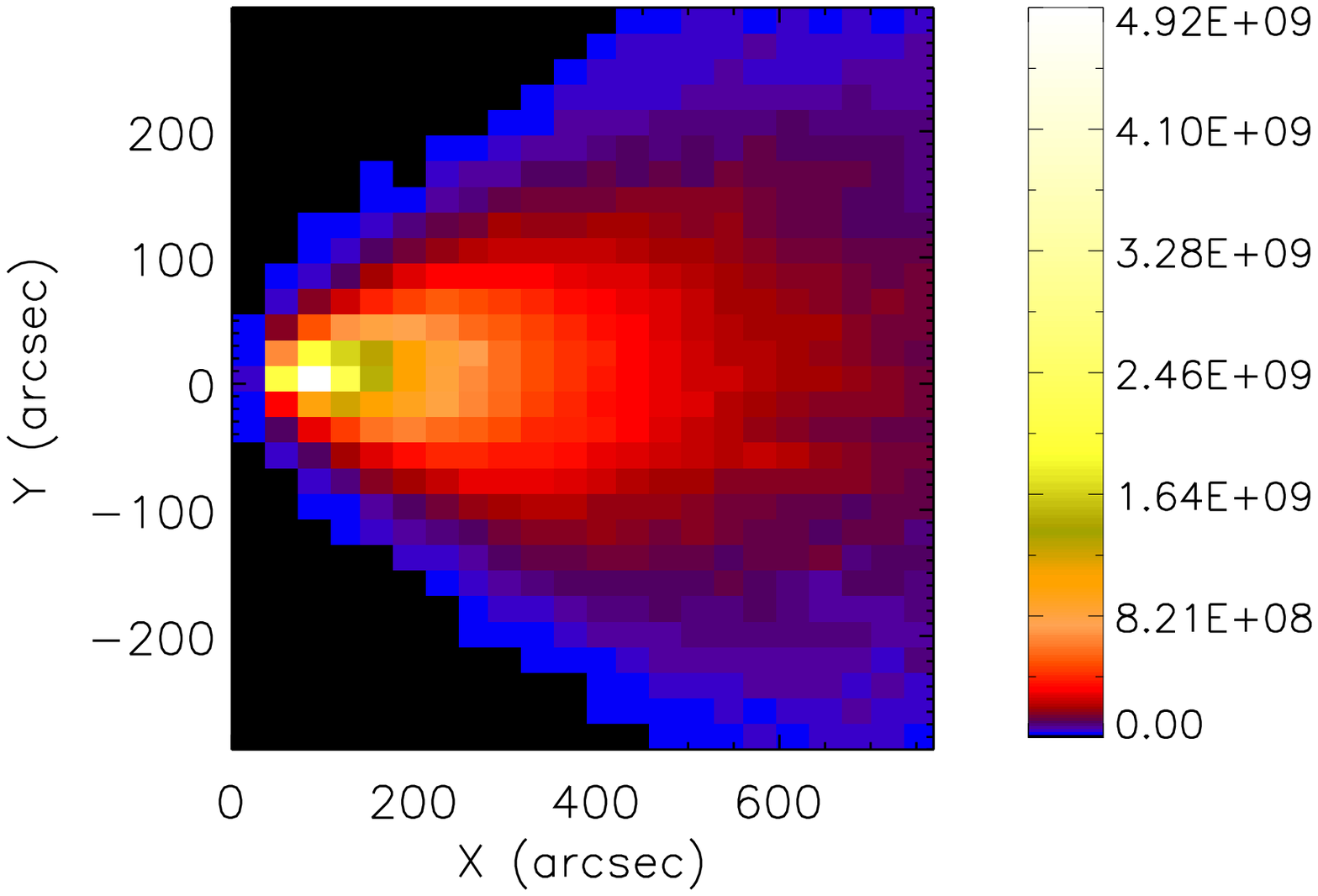} \\
  \end{tabular}
  \caption{\hi, \lya, intensity images reconstructed from the best simulation at $\rho_{obs}$=4.66~\rsun\, (left panel) 
  and at $\rho_{obs}$=5.72~\rsun\, (right panel)}
\end{figure*}

From the simulation runs that best fit the light curves we reconstruct the expected \hi\, \lya\, intensity 2D images
with the same procedure used to build the images of Figure~7.  The simulated images are shown in Figure~12. 
The images match quite well the observed data except that the comet nucleus is too bright in the simulation due to 
spatial blurring of the observational data which effectively reduces the spatial resolution.

As a further check on the goodness of fit we compute the variation of the size of the tail in the 
direction perpendicular to the comet path, that is along the UVCS slit, and compare
it with the observations. The tail size is defined as the $1/e$ distance from the intensity peak
determined from a gaussian fit of the total intensity profile along the slit. 
Figure~13 shows that the best fit model for the light curve also provides a good 
description of the tail size observed at $\rho_{obs}$=4.66~\rsun. 

\begin{figure*}[h]
  \centering
    \includegraphics[height=8.00cm,angle=90]{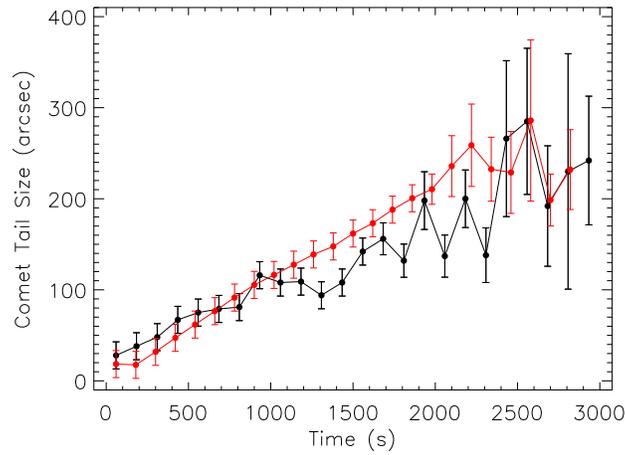}
  \caption{Simulated (red) and observed (black) comet tail size at $\rho_{obs}$=4.66~\rsun.} 
\end{figure*}

The gaussian fitting of simulated spectra also provides the variation of the line centroid and 
line width as the comet crosses the instrument slit.
We find that for the first two exposures, the simulated \hi\, \lya\, line is red--shifted by $\approx$+0.28~\AA\, 
and the line is very narrow ($\approx$0.12~\AA) because the signal comes from the cold first generation atoms 
moving with the nucleus away from the observer. 
The line width is in line with the observations (right panel of Figure~6), but the line centroid values are puzzling.
In first several exposures, there are large blue--shifts that cannot be due to instrumental effects (see also Section 2.2), 
whereas the simulation predicts spectral lines with small blue--shifts, $\approx$25--30 \kms,
as expected from ions moving with a 75~\kms\, solar wind at 39$^\circ$ to the line--of--sight.  
That is in line with the 0.1 to 0.2 \AA\/ shifts seen in most of the exposures.

We ran simulations for higher heights ($\rho_{obs}$=5.72~\rsun\, and 6.84~\rsun) 
with similar grids of parameters, except that the electron density grids are centered at 
the expected values at those heights.
The best fit between the observed and simulated light curves at $\rho_{obs}$=5.72~\rsun\, is shown in Figure~13. 
The coronal parameters are
$V_{w,0}$=75 \kms, $n_{e,0}=7.73\times 10^{3} cm^{-3}$ and $T_{k}=8\times 10^{5}$ K
and the outgassing rate is $\dot{N}_{0}=8.91\times 10^{27}s^{-1}$. 
As for the previous height, the best fit is obtained with an attenuation velocity of 3.5~\kms.

\begin{figure*}[h]
  \centering
    \includegraphics[height=8.00cm,angle=90]{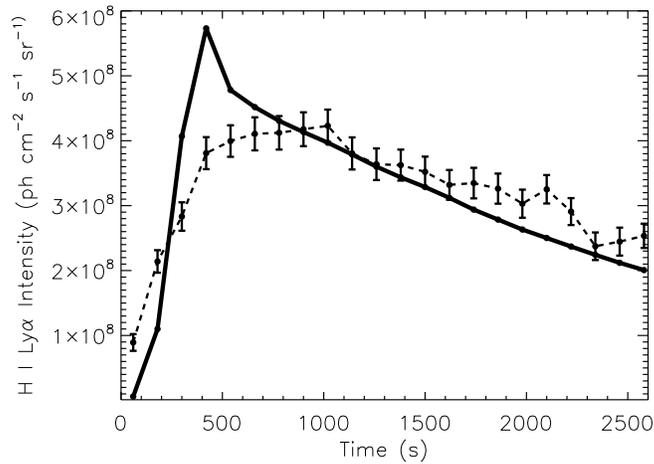}
  \caption{Best fitting of simulated (black) and observed (black dots with error bar) comet light curve 
  at $\rho_{obs}$=5.72~\rsun\, with the attenuation velocity of 3.5~\kms} 
\end{figure*}

\begin{figure*}[h!]
  \centering
  \begin{tabular}{cc}
   \hspace{0.0cm} \includegraphics[height=6.0cm]{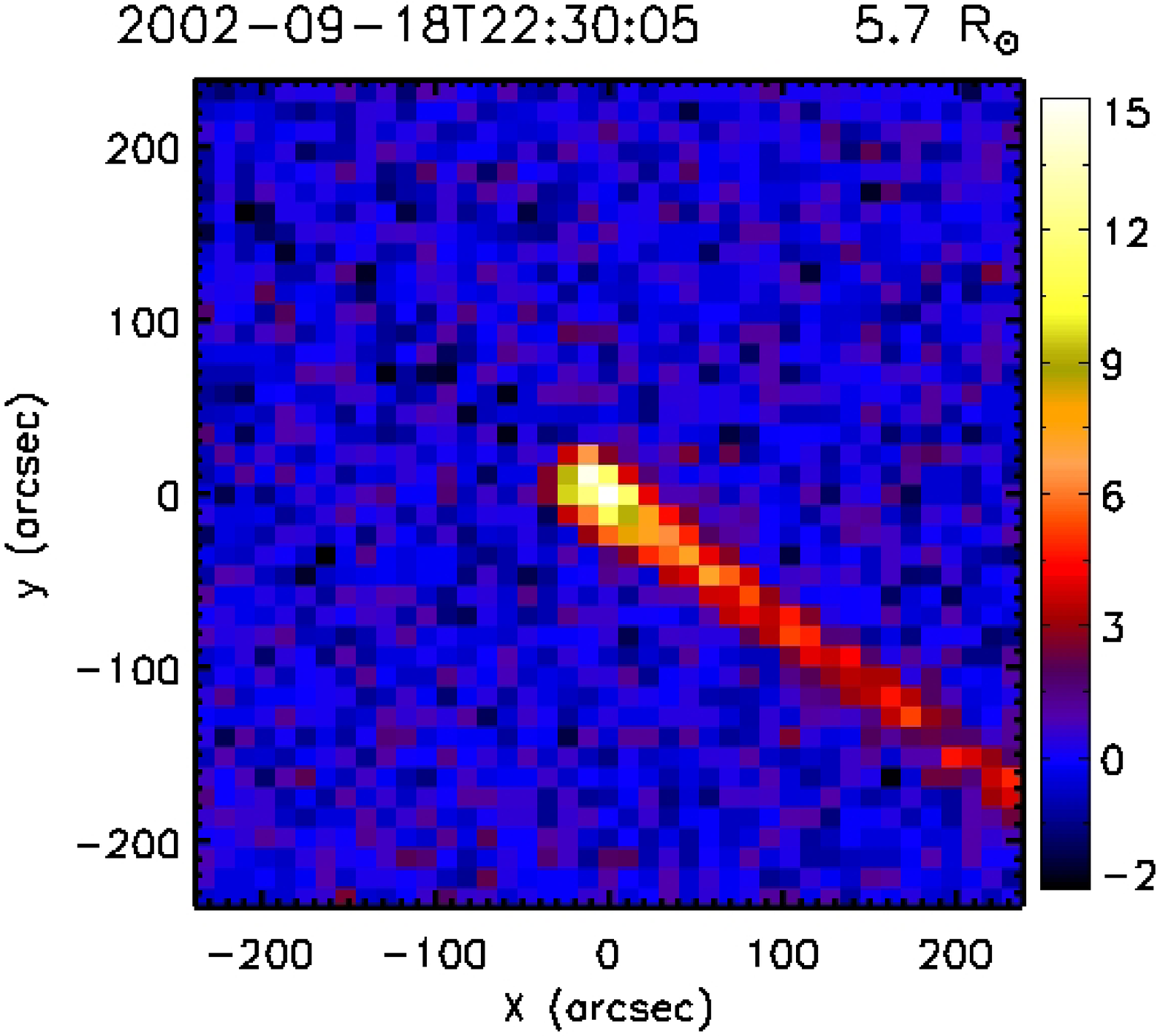} &
   \hspace{0.0cm} \includegraphics[height=6.0cm]{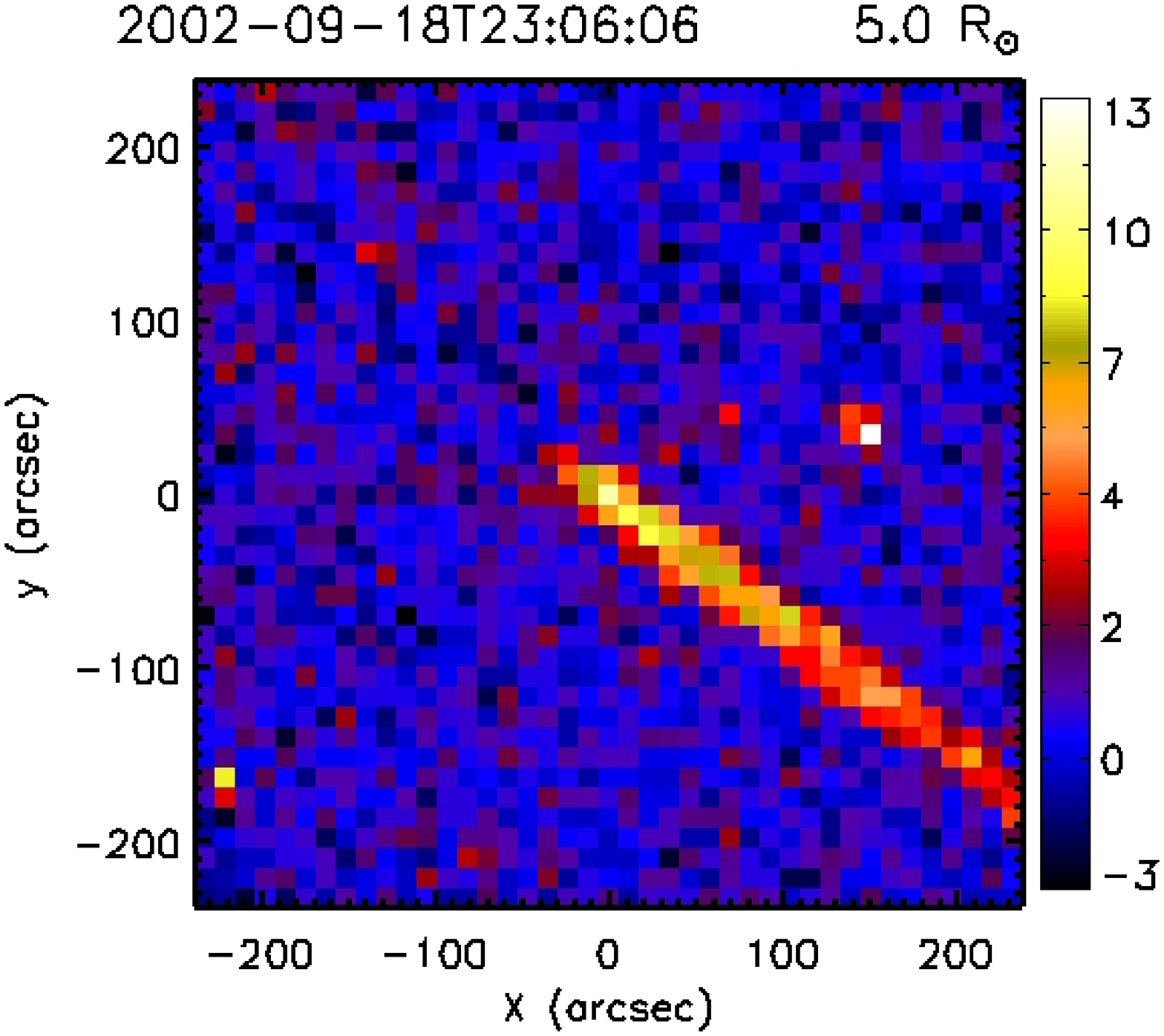} \\
  \end{tabular}
  \caption{LASCO C2 white light comet images at the time of UVCS observation at $\rho_{obs}$=5.72~\rsun, 
  at 22:30UT (left panel) and at 23:06UT (right panel)}
\end{figure*}

As we can see in the observed light curves at larger heights (Figure~5),  
the \hi\, \lya\, signal takes up to about 1000 s to reach the emission peak, then it slowly decreases.  
This trend is difficult to reproduce with our model which predicts a rapid rise when the comet nucleus 
reaches the slit, somewhat smoothed by the opacity and instrumental effects, followed by a rapid fading 
as it leaves the slit and a gradual fading after that. 
A possible explanation is the presence of many fragments traveling along the comet path, 
as supported by the LASCO C2 images of the comet archived with the Planetary Data System (Knight 2008). 
White light images have been processed by subtracting the median of four neighboring images 
to remove the background.  
Figure~15 shows the images at the time of UVCS observation at 5.72\rsun. 
We can see the clear difference between the two images taken with a time interval of 36 minutes: in the 
second one the nucleus is not well localized and the comet path seems to be a collection of many bright points.
However, this explanation seems at odds with the rapid rise seen at the lowest height.

Another explanation for the the slow fading of \hi\, \lya\, might be a tail of pyroxene grains. 
Kimura et al (2002) suggested that such grains could account for both the secondary peak
in LASCO brightness inside 7 \rsun and also for the increased \lya\, emission at small heights \citep{uzz01}
if pyroxene sublimates and acts to neutralize protons in the corona.  Bemporad et al 2005 attributed a
very slowly decaying \lya\ signal in comet C/2001 C2 to sublimation of dust grains in the comet tail and 
subsequent charge transfer interactions.
A 0.1 $\mu$m pyroxene grain is expected to survive for about 1000 seconds at about 5 \rsun\, 
before it sublimates (Kimura et al 2002).

\begin{figure*}[h!]
  \center
    \includegraphics[height=8.0cm,angle=90]{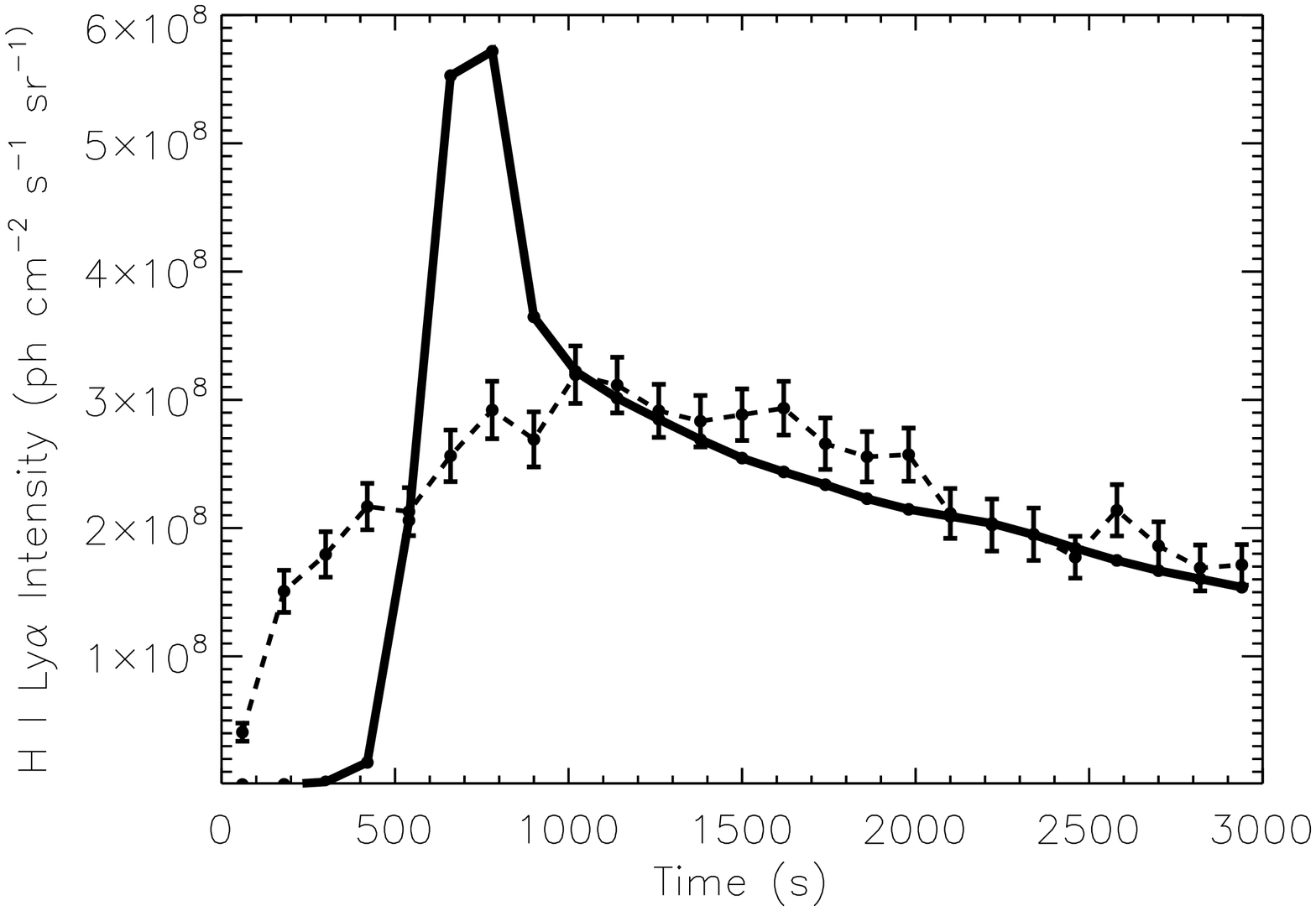}
  \caption{Best tail fitting of simulated (black) and observed (black dots with error bar) comet light curve at $\rho_{obs}$=6.84~\rsun.}
\end{figure*}

None of the models could match the light curve at the third height ($\rho_{obs}$=6.84~\rsun ; see Figure~16).
One problem is that the relatively low speed of the comet at this height leads to a bright spike due to
the first generation neutrals.  The second problem is the very slow rise.  We discuss model limitations
that could cause these problems, along with the red-- and blue--shifts seen at different positions along
the slit at lower heights, in the next section.

\section{Discussion}

The Monte Carlo model allows us to interpret UV and optical light curves in terms of the 
coronal and cometary parameters and the ion and dust contributions.  
A more complete model will be needed to analyze the interaction between the ions produced 
by the comet and the ambient solar wind and coronal magnetic field, 
but we can make an initial assessment, discuss and summarize our results.

\subsection{Coronal density, temperature and solar wind}

At the two lower observed heights, the model can reproduce the 
\hi\, \lya\, comet light curves and the size of the comet tail perpendicular to comet path. 
Table~4 gives the results of the model fits; we point out that the determined parameters are related to
the actual heliocentric distance, $r$, whereas $\rho_{obs}$ is just the projection on the plane of the sky.
The coronal electron densities are compared with the values derived
for the solar maximum coronal current sheet and coronal hole by 
\citet{guh06}, which can be taken as the lower and upper limits to coronal density (Figure~17). 
The uncertainty on the density is given mainly by the grid of values used for the simulation.
The best fit wind velocity is 75 \kms\, at both heights, though at r=6.97~\rsun\, a value of 100~\kms\, 
is also acceptable, as expected from the wind acceleration.
Higher wind velocities do not predict persistent H I comet tails because of Doppler dimming.
One way to get acceptable light curves with higher wind speeds would be to simulate a 
number of well separated fragments, but that is beyond the scope of this paper. 
The proton kinetic temperature is typically $8 \times 10^5$K, as confirmed 
by the size of the comet tail in the direction perpendicular to its path and from the line widths.
 
\begin{table}[h]
\caption{Best model parameters} 
\centerline{
\begin{tabular}{|c|c||c|c|c||c|c|}
\hline
$r$	& $\rho_{obs}$	& $n_{e,0}$ 		& $V_{w,0}$	& $T_{k}$ 			& $\dot{N}_{0}$		& $V_{att}$	\\
\rsun	&\rsun		& $\rm cm^{-3}$		& \kms	& K				& $\rm s^{-1}$		& \kms	\\ 
\hline                                      
5.99	& 4.66		& $1.23 \times 10^{4}$ 	& 75 		& $8.0 \times 10^{5}$ 	& $1.12 \times 10^{28}$	& 3.5	\\
6.97	& 5.72			& $7.73 \times 10^{3}$	& 75 		& $8.0 \times 10^{5}$	& $8.91 \times 10^{27}$	& 3.5	\\
\hline
\end{tabular}
}
\end{table}

\begin{figure*}[h]
  \centering
    \includegraphics[height=8.00cm,angle=90]{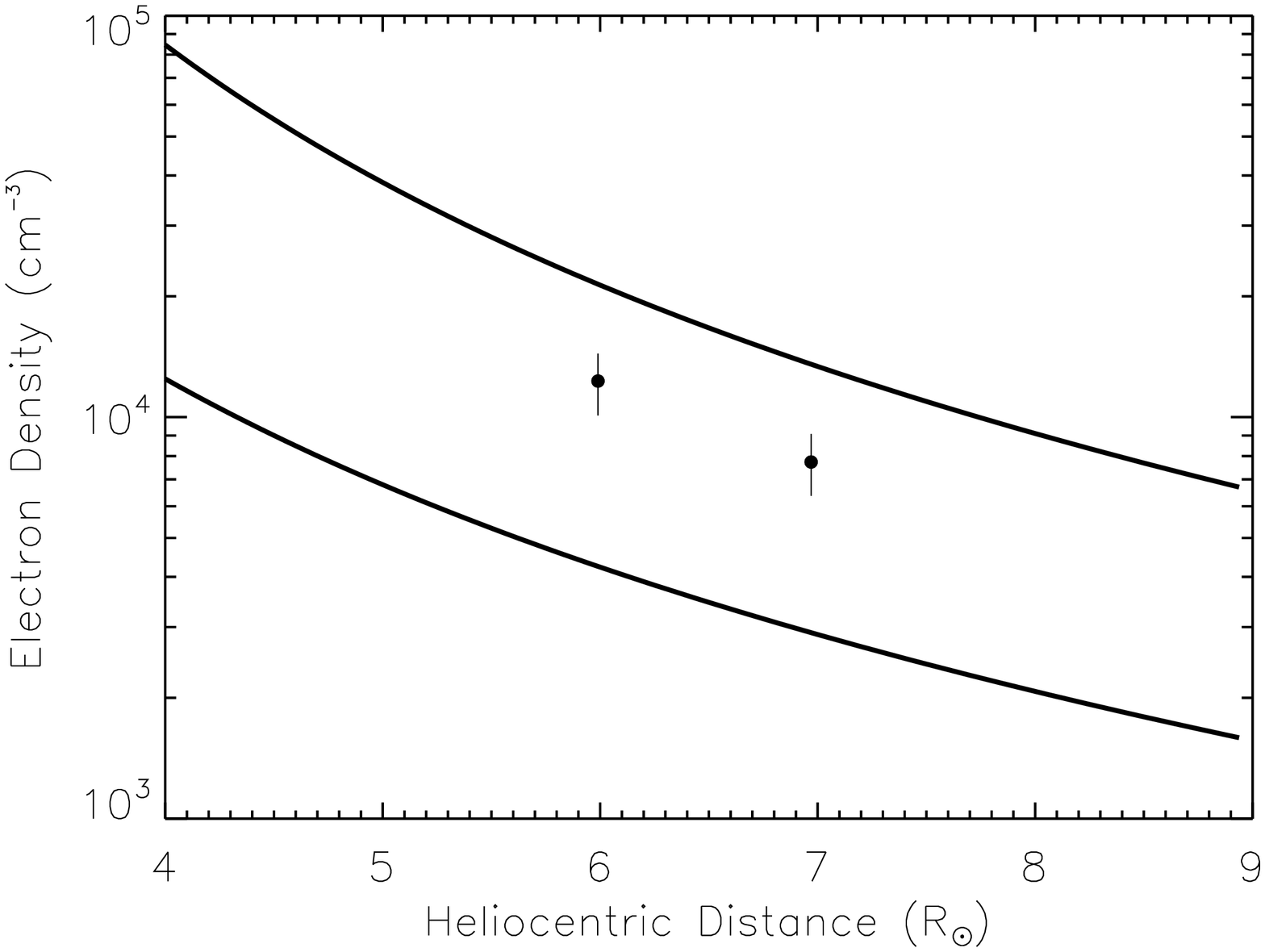}
  \caption{Coronal electron density at the heliocentric distance of UVCS comet observation (full dots with error bars). 
  The two curves shows the electron density for current sheet and coronal hole at solar maximum 
  measured by \citet{guh06}} 
\end{figure*}

\subsection{Outgassing rates and size of nucleus}

As the comet approaches the Sun, it experiences a rapid increase in illumination, and
most of the energy of the solar radiation goes into the sublimation of water.
The expected water outgassing rate, $\dot{N}_{\small \water}$ in molecules per second,
can be estimated by assuming a balance between the energy supplied by the solar radiation over the cometary active surface, $S_{act}$,
and the energy required to sublimate the quantity of ice corresponding to $\dot{N}_{\small \water}$.
Therefore, the energy balance can be written as
\begin{equation}
F_{h}~S_{act}~\approx~\dot{N}_{\small \water}\frac{L}{N_A} \frac{1}{1-A}
\end{equation}
where
$F_{h}=F_{\odot} / h^2$ is the solar flux, in erg cm$^{-2}$ s$^{-1}$,
scaled to the cometary heliocentric distance in AU, $h=(r / 215$ \rsun$)$,
$F_{\odot} = 1.37\cdot10^6$ erg cm$^{-2}$ s$^{-1}$ is the solar flux at 1AU,
$L = 4.81 \cdot 10^{11}$ erg mol$^{-1}$ is the latent heat sublimation of ice,
$N_A=6.022 \times 10^{23}$ molecules mol$^{-1}$ is Avogadro's number and
$A = 0.06$ is the cometary albedo.
In a spherical approximation, the active surface exposed to the solar radiation is
$S_{act}=\pi r_{com}^{2}$, where $r_{com}$ is the equivalent cometary nucleus radius.
From Equation~12, $\dot{N}_{\small \water}$ can be estimated as
\begin{equation}
\dot{N}_{\small \water}~\approx ~\pi  F_{\odot} \frac{r_{com}^2}{h^2}  N_A \frac{1-A}{L}
\end{equation}

The best fit models at the two lower heights provide estimates of the neutral hydrogen outgassing rate. 
By assuming that all the observed H originates in the dissociation of water we can determine
water production rate as $Q_{\small \water} = 18 \cdot \dot{N}_{\small \water} \cdot U$,
where $\dot{N}_{\small \water}=\dot{N}_H/2$ and $U= 1.66 \times 10^{-27}$ kg is the Atomic Mass Unit.
The values are reported in Table~5, with the collection of published results obtained at different heliocentric 
distances from sungrazers observed by UVCS. 

In a spherical approximation, from the values of the water production rate, $\dot{N}_{\small \water}$, 
by using Equation~13, we can derive an estimate of the comet equivalent radius, $r_{com}$, as:
\begin{equation}
r_{com}=h\sqrt{\frac{1}{\pi}\frac{\dot{N}_{\small \water}L}{F_{\odot}(1-A)N_A}}
\end{equation}
where all the terms are described in Section~3.1.  Comet Lovejoy (C/2011 W3) was by far the largest 
sungrazing comet observed in recent years, with a diameter of about 400 m \citep{mccauley13}.

If the comet has fragmented, as suggested by Figure~15, the inferred radii shown in Table~5 are upper limits.
The sizes of the nucleus shown Table 5 do not show any increase in the effective diameter between the crossing
at 6.97 \rsun\, and the crossing at 5.99 \rsun\, as seen between 5.88 \rsun\, and 4.68 \rsun\, for comet C/2000 C6 
and interpreted as an indication of fragmentation by \cite{uzz01}. 
However, if the nucleus were really a 9.4 m diameter sphere at 6.97 \rsun , 
it would shrink by about 1 m at the outgassing rate shown in Table~5 by the time it reached 5.99 \rsun. 
The smaller change in apparent diameter might suggest some partial fragmentation, but the inferred diameters
are uncertain at at least the 10\% level even in the absence of fragmentation.

\begin{table}[h]
\caption {Previously published sungrazing outgassing rates.}
\centerline{
\begin{tabular}{|l|c|c|c|c|}
\hline
Comet           	& r			& $\dot{N}_H$			& $Q_{\small \water}$  	& $r_{com}$		\\
Name        	   	& [\rsun]		& [$10^{28}\rm s^{-1}$]& [$\rm kg~s^{-1}$] 	& [m] 		\\
\hline
C/2002 S2		& 5.99		& 1.12			& 167				& 9.0		\\
			& 6.97		& 0.89			& 133				& 9.4		\\
C/1996 Y1$^{a}$	& 6.80		& 0.13			& 20.0			& 3.4		\\
C/2000 C6$^{b}$	& 3.88		& 0.71				& 71.8			& 3.0		\\
			& 4.68		& 1.35			& 140.0			& 5.7		\\
			& 5.88		& 0.33			& 34.6			& 3.4		\\
			& 6.47		& 0.13			& 10.5			& 2.5		\\
C/2001 C2$^{c}$	& 4.98$^{e}$	& 0.59			& 58.9			& 7.8		\\
			& 4.98$^{f}$	& 0.29			& 28.5			& 5.4		\\
			& 3.60		& 8.20			& 820				& 20.3	\\
C/2003 K7$^{d}$	& 3.37		& 40 -- 170			& 6000--25000		& 29--60	\\
\hline
\end{tabular}
}
\end{table}		
\noindent
$a.$ \cite{ray98B}\\ 
$b.$ \cite{uzz01}\\
$c.$ \cite{bem05}\\ 
$d.$ \cite{cia10}\\
$e.$ Main fragment\\ 
$f.$ Subfragment\\

\subsection{UV and Optical light curves}

The optical fluxes measured by LASCO include both solar continuum scattered by dust grains and 
resonantly scattered emissions, prominently Na I \citep{kni10}.
The sudden drop in brightness inside 11~\rsun\, with no counterpart in \hi\, \lya\, (Figure~3) 
indicates that it corresponds to a sudden change in the dust destruction rate, 
for instance sublimation of the olivine grains when their temperature exceeds about 1500 K \citep{kim02}.
The sudden drop also implies a change in the radiation due to Na I and other atomic or molecular species. 
It may be that the temperature in the coma becomes high enough to dissociate the molecules and ionize Na I.
\citet{kni10} find an $r^{-4}$ brightening inside 24~\rsun\,,
which is attributed to the $r^{-2}$ dilution factor and $r^{-2}$ outgassing rate.

The optical and UV light curves in Figure~3 are strikingly different.
The rapid increase in Ly$\alpha$ brightness shows that the outgassing rate is rising rapidly even as
the optical brightness drops by an order of magnitude.
Sungrazing comets typically brighten as $r^{-3.8}$ in the optical as they move from 24 \rsun\/ to 16 \rsun,
reach a peak near 12 \rsun\/ \citep{bie02,kni10} and fade by two or three magnitudes as they move in
to around 6 \rsun.
Some brighten again when they reach about 5 to 7 \rsun, C/2002 S2 being an extreme case.
The rapid fading inside 12 \rsun\/ is usually attributed to sublimation of dust grains 
\citep{bie02,uzz01,kim02,bem05,kni10} and the brightening at smaller radii is attributed either to an increase in the
outgassing rate as the comet nucleus fragments and exposes more surface area to
sunlight \citep{bie02,uzz01} or to crystalline pyroxenes, a more refractory
form of dust \citep{kim02} .
The dust is probably composed primarily of silicates rather than carbonaceous compound \citep{kim02}.
\citet{cia10} found a Si:C ratio of about 10 in the sublimated dust of another Kreutz sungrazer,
and carbonaceous grains would rapidly vaporize at the temperatures expected this close to the Sun.

The optical fluxes measured by LASCO include both solar continuum scattered by dust
grains and resonantly scattered features such as Na I as indicated by a difference of
up to 1.5 magnitudes between observations with the orange and clear filters \citep{kni10, lam13}.
The sudden drop in brightness suggests a change in the radiation due to Na I and other atomic or molecular
species as well as destruction of dust grains.  It may be that the temperature in the coma becomes 
high enough to dissociate the molecules and ionize Na I.
\citet{bie02} found that the emission line contribution is small by the time a sungrazer reaches 7 \rsun,
while Figure 2 of \cite{kni10}, which shows C2 magnitudes for this comet systematically brighter than C3, 
suggests that the line contribution could be up to 30\% of the C2 brightness.

\subsection{Disturbance of the corona by the comet}

Thus far we have treated the comet as a test particle, in that it has no effect
of the solar wind it encounters.  However, the comet will significantly perturb the wind
in ways that can affect both the light curves and the Doppler shifts.
As discussed by \cite{galeev85} and \cite{gom96}, material from the
comet mass loads the solar wind, and if the outgassing rate is large
enough it can create a bow shock.  The wind speed obtained by fitting the light
curve should be the speed of the wind that has been slowed by the interaction.

We can estimate the importance of the dynamical effects of the interaction in two ways.  
First, we can estimate the size of the bow shock, and second we can compare the
mass lost from the comet with the coronal mass in the interaction region.  

The standoff distance, $r_{st}$, from the comet nucleus is
obtained by equating the ram pressure of the outflowing cometary gas to the
ram pressure of the gas it encounters,

\begin{equation}
\frac{\dot{N}_H}{4 \pi r_{st}^2 V_{out}}\mu_{com} V_{out}^2~=~ n_{cor} \mu_{cor} V_{rel}^2
\end{equation}

\noindent
where $V_{rel} \approx V_{r}-V_{w}$ is the comet speed relative to the solar wind assumed to be radial,
$\mu_{cor}$ = $1.8 \times 10^{-24}$ g is the coronal mass per H nucleus for a wind with 5\% helium,
and $\mu_{com}$ = $1.49 \times 10^{-23}$ g is the mass per H nucleus for water.
At the lowest observed comet height, r=5.99\rsun, where the comet velocity is $V_{r}$=227 $\rm km~s^{-1}$,
the standoff distance can be evaluated for the coronal and cometary parameters in Table~4.
For a coronal density $n_e = 1.2 \times 10^4$ $\rm cm^{-3}$,
a wind speed $V_{w} \approx 100$ $\rm km~s^{-1}$
an outgassing rate $\dot{N}_H = 1.1 \times 10^{28}$ $\rm s^{-1}$ and
an outgassing speed $V_{out} \approx 3.5$ $\rm km~s^{-1}$ from the optical depth analysis,
the standoff distance is only 140 km.
The lateral extent of the bow shock size around twice this size, but even so, it is modest
compared to the mean free path of the neutrals.
\citet{gom96} presented a detailed model of a comet bow shock driven by outgassing hydrogen.
They show that many of the neutrals charge exchange outside the bow shock, so that the mass--loaded flow
slows down over a large region ahead of the shock.  For the parameters relevant here,
we expect a Mach number of around 3, so the compression and heating will be modest.

A second estimate of the importance of the interaction is to compare the mass lost from the comet with the
mass of the solar wind in the interaction volume.
The comet mass loss rate is $\dot{N}_H \mu_{com}$.
The outgassed atoms interact with the ambient corona through the lifetime before ionization, $\tau_d$,
or about 950 s for a density of $1.2 \times 10^4 \rm cm^{-3}$ and a relative speed of
$V_{rel}~\sim 330~\rm km~s^{-1}$. 
The corresponding interaction volume is a cylinder whose length is $V_{rel} \times \tau_{d}$,
or $3.3 \times 10^5$ km, and radius is the distance a particle travels during the interaction time,
$V_{exp} \times \tau_{d}$, where $V_{exp}$ is the expansion speed given by the rate of growth
of the width of the tail seen in Figure~13.  For $V_{exp}$ = 75 \kms\, the
cylinder radius is about 72,000 km.  Then the coronal mass in the interaction cylinder is about  
$10^7$ kg, compared with an outgassed mass of
about $4.8 \times 10^5$ kg in the time $\tau_{ex}$ for $\dot{N}_H$ = $1.1 \times 10^{28}~\rm s^{-1}$.
The actual size of the interaction region is smaller, because the oxygen atoms that account for
most of the mass are moving more slowly, and no single scale length applies to all the interactions.
Thus the comet significantly perturbs the solar wind with which it interacts, but does not dominate
the flow.  The sense of this perturbation is mostly a motion along the comet's path, which in this case
is toward the Sun and away from the Earth.

One important effect of the disturbance in the solar wind velocity is that the bow shock created 
by the cometary plasma produces a region of heated coronal gas in the stagnation region at the tip 
of the bow shock which moves with the comet. 
Atoms that undergo charge transfer in that region produce a cloud of neutrals centered on the comet
that expands at approximately the relative speed of the comet and the wind. 
Charge transfer in the flanks of the bow shock and in the region outside the bow shock that is 
compressed and accelerated by the interaction \citep{gom96} will produce neutrals expanding more
slowly about centroid velocities intermediate between the comet speed and the wind speed. 
Some of the neutrals produced in these heated, accelerated regions can move upstream faster than the comet. 
They account for the somewhat gradual rise of the light curve at 6.97 \rsun\, and the even slower rise 
at 8.02 \rsun .  

In addition, neutrals created near the
stagnation region near the tip of the bow shock have a velocity centroid
equal to that of the comet, so in the case of C/2002 S2 they are moving
away from the Earth and might account for the red--shifts seen on the
lower sides of the velocity images in Figure~9.  However, it is not obvious
why the red--shifted material would lie along one side of the comet trajectory,
because to first order the bow shock is symmetric.  
The symmetry is broken by the fact that the comet speed is not antiparallel to the solar wind speed.
In addition, the magnetic field can distort the bow shock in the same way
that it affects the shape of the heliospheric termination shock \citep{lallement05, opher09}.
A detailed model would be needed to determine whether this distortion
could produce spatially separated red-- and blue--shifted regions.

\subsection{Pickup Ion Behavior}

The pattern of blue--shifted emission to the north and red--shifted
emission to the south seen in Figure 9 is quite remarkable.  It has
not been seen in other comets observed by UVCS, but that could be due
to the viewing geometry or the lower signal--to--noise ratio of most of
the other observations.  

At first glance, the blue--shift can be explained as the line--of--sight component of the
solar wind speed.  For observed blue--shifts up to 100 $\rm km~s^{-1}$
and the phase angles listed in Table 2, the solar wind speed would
be of order 150 to 200 $\rm km~s^{-1}$ if the wind is flowing radially.
Those speeds agree reasonably well with streamer velocities at those
heights \citep{str02, frazin03, abb10B}.  However, as discussed in connection
with the Monte Carlo models, speeds that large imply severe Doppler dimming
and require correspondingly large outgassing rates that would be hard to reconcile 
with the lack of \lyb\, emission and the lack of a narrow spike from first
generation hydrogen atoms.  Models that match the light curves reasonably well
predict blue--shifts below about 60 \kms.

The red--shift and its southward displacement must be related to the line--of--sight
component of the comet's velocity away from the Earth.  However, in detail the
red--shift and displacement are complex to model.
The most likely explanation for the red--shift along one edge of the comet's
\hi\, \lya\, tail is related to the interaction of the cometary H atoms with
the solar magnetic field.  When a neutral moving with the comet is ionized
in the coronal magnetic field, it behaves as a pickup ion, with its velocity
component parallel to the field giving a flow along the field, whereas the
perpendicular component becomes the gyro velocity around the field.  
The resulting ring beam in velocity space is unstable, and it rapidly evolves 
to a more isotropic bispherical shell in velocity space \citep{williamszank}.  

The resulting ions move along the field,
as is seen in the striations observed in oxygen ions in Comet Lovejoy C/2011 W3
\citep{downs13, raymond14}.  Because Comet C/2002 S2 is moving away from the
Earth, $V_{\|}$ gives a red--shift.  As seen in Comet Lovejoy, the gas can move to
either side of the comet trajectory depending on the orientation of the
magnetic field.  Based on the observed red--shifts and the dominance of red--shift
to one side of the comet tail, the line--of--sight component of the parallel
speed is of order 80 \kms, whereas the plane--of--the--sky component must
be comparable to the thermal speed that causes the tail to expand, or 100 \kms.
In order for these pickup ions to become visible in \lya , some of them
must undergo a second charge transfer event to produce a population of
neutrals with the pickup ion velocity distribution (third generation neutrals).
The relatively high outgassing rate of Comet C/2002 S2 makes this second charge 
transfer more likely than in the smaller sungrazers observed by UVCS.
If the plane--of--the--sky component of the solar wind speed tends toward the
north, while the plane--of--the--sky component $V_{\|}$ tends toward the south,
this explains the separation of red-- and blue--shifted emission seen in Figure~9. 

In earlier studies of sungrazing comets with UVCS, we were able to
estimate the coronal density and temperature from the \hi\, \lya\, decay
time and line width.  The blue--shift adds a diagnostic for the solar wind
speed, and the red--shift and the separation of the red-- and blue--shifted
components offer a means to assess the magnetic field direction in 3
dimensions.  However, the modeling needed to extract that information is
a step beyond the models presented here.

\section{Summary}

We report the UVCS/SOHO and LASCO/SOHO observations of sungrazing Comet C/2002 S2 at three
heights, and we describe Monte Carlo simulations used to interpret the data
in terms of the comet outgassing rate and to probe the coronal density, 
temperature and flow speed along its path.  The Monte Carlo simulations
are able to match the light curves and the expansion of the emission
region along the UVCS slit at the lower two heights, but they do not
match the light curve at the upper height.
This comet is unlike others observed by UVCS in that it shows distinct red-- 
and blue--shifts below and above the center of the images, and the Monte
Carlo simulations do not predict such behavior.

There are two effects that were not included in the Monte Carlo models because they
can be neglected for comets with small outgassing rates, but that are probably
important for Comet C/2002 S2 and may account for the discrepancies between
models and observations.  First, the mass loading of the solar wind by cometary
material slows the flow in the comet's frame of reference, compresses it and
may cause a bow shock.  Second, the protons produced when cometary neutrals
are ionized by charge transfer, collisions with electrons or photoionization
become pickup ions, and if those ions charge transfer with other neutrals they
create a third generation population of neutrals with a distinctive mean velocity
and velocity width.  We plan to include the first effect in the next generation
of Monte Carlo models.  The second effect will be deferred to the more distant future
because it is more complex, and it involves additional free parameters.

Because of the good capability of the Monte Carlo model to describe the \hi\, \lya\, emission 
from small sungrazers at low heliocentric distances, we plan to apply the method to other comets
detected by UVCS, moreover we will adapt the code to compute the expected UV images, 
for a grid of coronal and cometary parameters, which shall be obtained by the coronagraphs aboard 
future space mission, such as the Solar Orbiter.

\acknowledgements
We are indebted to Brian Marsden, who performed the orbit calculations that made these observations possible.  
SG and JR thank the International Space Science Institute (Bern, Switzerland) for the opportunity 
to discuss this work within the International Study Team programme.
SOHO is a project of international cooperation between ESA and NASA.
UVCS is a joint project of NASA, Italian Space Agency (ASI),
and the Swiss Funding Agencies.
LASCO was built by a consortium of the Naval Research Laboratory, USA, the Laboratoire 
d'Astrophysique de Marseille (formerly Laboratoire d'Astronomie Spatiale), France, 
the Max--Planck--Institut f\"ur Sonnensystemforschung (formerly Max Planck Institute 
f\"ur Aeronomie), Germany, and the School of Physics and Astronomy, University of 
Birmingham, UK. 
This work was supported by NASA grant NAG5--12814.

{\it Facilities:} \facility{SOHO (UVCS, LASCO)}


\bibliographystyle{apj}

\end{document}